\documentclass[11pt,preprint2,longabstract]{aastex}

\shortauthors{Lee et al.}
\shorttitle{Increasing SFHs of high-$z$ LBGs}

\begin{document}

\title{Steadily Increasing Star Formation Rates in Galaxies Observed at $3 \lesssim z \lesssim 5$ 
in the CANDELS/GOODS-S Field}

\author{Seong-Kook Lee}
\affil{Center for the Exploration of the Origin of the Universe, Department of Physics and Astronomy, Seoul National University, Seoul, Korea}
\email{sklee@astro.snu.ac.kr}

\author{Henry C. Ferguson}
\affil{Space Telescope Science Institute, 3700 San Martin Drive, Baltimore, MD 21218}

\author{Rachel S. Somerville}
\affil{Department of Physics and Astronomy, Rutgers, The State University of New Jersey, 136 Frelinghuysen Road, Piscataway, NJ 08854}

\author{Mauro Giavalisco}
\affil{Astronomy Department, University of Massachusetts, 710 North Pleasant Street, Amherst, MA 01003}

\author{Tommy Wiklind}
\affil{ESO/Joint ALMA Observatory, 3107 Alonso de Cordova, Santiago, Chile}

\and

\author{Tomas Dahlen}
\affil{Space Telescope Science Institute, 3700 San Martin Drive, Baltimore, MD 21218}

\begin{abstract}

We investigate the star formation histories (SFHs) of high redshift ($3 \la z \la 5$)
star-forming galaxies selected based on their rest-frame ultraviolet (UV) colors in
the CANDELS/GOODS-S field. By comparing the
results from the spectral-energy-distribution-fitting analysis with two different
assumptions about the SFHs --- i.e., exponentially declining SFHs as well as increasing
ones, we conclude that the SFHs of high-redshift star-forming galaxies increase with
time rather than exponentially decline. We also examine the correlations between 
the star formation rates (SFRs) and the stellar masses.
When the galaxies are fit with rising SFRs, we find that the trend seen in
the data qualitatively matches the expectations from a semi-analytic model of galaxy formation.
The mean specific SFR is shown to increase with redshift, also in 
agreement with the theoretical prediction. 
From the derived tight correlation between stellar masses and SFRs, we
derive the mean SFH of star-forming galaxies in the redshift range of $3 \leq z \leq 5$,
which shows a steep power-law (with power $\alpha = 5.85$) increase with time. We also
investigate the formation timescales and the mean stellar population ages of these 
star-forming galaxies. Our analysis reveals that UV-selected
star-forming galaxies have a broad range of the formation redshift.
The derived stellar masses and the stellar population ages show positive correlation in a
sense that more massive galaxies are on average older, but with significant scatter. This large 
scatter implies that the galaxies' mass is not
the only factor which affects the growth or star formation of high-redshift galaxies. 

\end{abstract}


\section{Introduction}
Knowledge of the physical properties --- such as stellar masses, star-formation rates (SFRs), and stellar population ages --- of galaxies is indispensable for understanding the evolution and formation of galaxies.
Thanks to the last decade's boom in the panchromatic observation of remote galaxies, which has benefited from the development of powerful facilities, including the $Hubble$ $Space$ $Telescope$ ($HST$), a great advance has been made in the study of high-redshift galaxies as well as in our understanding of galaxy evolution. This advance in multi-wavelength studies of high-redshift galaxies is expected to accelerate in near
future with the arrival of powerful space telescopes, such as $James$ $Webb$ $Space$ $Telescope$, as well as large ground facilities, including the Giant Magellan Telescope and the Thirty Meter Telescope.

However, constraining the star-formation histories (SFHs) of galaxies from observation is not an easy task, even with these panchromatic data over a wide wavelength range. The difficulty comes from several factors : for example, the degenerate effects between
SFHs and other properties of galaxies --- such as dust extinction, metallicity, and redshift --- on the
overall spectral energy distributions (SEDs) of galaxies, and the fact 
that the observed SEDs can be easily dominated by the light from massive, 
young stars, readily concealing the old stellar populations.

One way to derive physical inferences from the multi-band photometry is to compare 
the distribution of galaxy colors and magnitudes to the predictions of theoretical 
models of galaxy formation \citep[e.g.][]{som01,som08,som12,idz04,nag05,men06,nig06,fon09}.

Another way is to derive physical parameters, such as stellar masses, SFRs, 
metallicities, and constraints on SFHs, by fitting the SED of each individual galaxy.
This method is very useful in studying galaxies' physical parameters or stellar population
properties of unresolved galaxies, and has been used in analyzing various galaxy populations over a
wide redshift range \citep[][just to name a few]{saw98,pap01,sha01,sal05,guo12}.

In this SED-fitting analysis, we should inevitably make assumptions on several properties 
of galaxies --- including their stellar initial mass function (IMF), chemical composition 
and its evolution, dust-attenuation law, and SFHs of galaxies. The assumption on each of 
these properties can affect the SED-fitting results, and has been studied by several 
authors --- for example, the effects of IMF are studied by \citet{pap01} and \citet{con09}, 
the effects of stellar evolution 
model are investigated by \citet{mar06}, and the effects of metallicity evolution are 
studied by \citet{con09}. 
Detailed review on this issue can be found in \citet{con13}

Among these assumptions, the effects of assumed (forms of) SFHs have 
been studied by \citet[][hereafter L10]{lee10}. In L10, comparing with the results from 
\citet[][hereafter L09]{lee09}, we have extensively analyzed the effects of assumed SFHs 
on the SED-fitting results of $3 < z < 6$ Lyman-break galaxies (LBGs) by analyzing mock 
LBGs from semi-analytic models (SAMs) of galaxy formation. 
This analysis has revealed that the assumptions about SFHs can 
significantly bias the inferences about stellar-population parameters: particularly 
ages and SFRs. 
Also, the results of L09 and L10 --- combined with the prediction from SAMs of galaxy 
formation (represented in L10) --- suggest that the SFHs of star-forming galaxies at 
this redshift range ($3 < z < 6$) increase with 
time --- in contrast to the assumption of declining SFHs as done in some previous works.

For observed galaxies, \citet{mar10} show that the SEDs of $z \sim 2$
($BzKs$-selected) star-forming galaxies are fitted better with the exponentially
increasing SFHs rather than exponentially decreasing SFHs.
\citet{pap11} have analyzed the evolution of average SFRs of
high-redshift galaxies, by studying the high-redshift galaxy samples with
constant comoving number density, n=$2 \times 10^{-4}$ Mpc$^{-3}$, from $z=8$
to $z=3$. By following the galaxy samples with same comoving number density,
they can study the evolution of SFRs and stellar masses of galaxies, which
may be connected as the predecessors and their descendants.
From this study, they show that the average SFR of high-redshift galaxies
increases as a power law with decreasing redshift (with best-fit power
$\alpha = 1.7$).

Motivated by the evidence favouring increasing SFHs, we analyze 
the SEDs of $3 \leq z \leq 5$ LBGs observed in the southern field of the Great 
Observatories Origins Deep Survey (GOODS-S). 
Specifically, we focus on the SFHs of these observed galaxies, and examine if 
assuming the increasing SFHs would be more appropriate, providing better constraints 
on the stellar population properties of the observed high-redshift galaxies
than assuming the declining SFHs in SED-fitting analysis.

The relatively tight correlation between stellar masses and SFRs for star-forming 
galaxies at redshifts up to $z \sim 2$ has come to be known as the star-forming 
main sequence \citep{dad07,elb07,noe07,pan09}. 
Over a wide redshift range, this correlation remains tight with near unity slope 
while the normalization evolves quickly with redshift. 
This tight correlation of galaxies' star-formation activities with their already-formed 
stellar masses, the nearly invariant slope of this correlation, and the evolution of 
its normalization in a broad redshift range must impose important constraints on the 
formation histories of these star-forming galaxies (for example, see \citet{ren09} 
and \citet{saw12}). 
In this work, we investigate this correlation and its evolution in the redshift 
range between $z \sim 5$ and $z \sim 3$, and derive meaningful constraints on the 
formation histories of the star-forming galaxies at this redshift range.

Regarding the SFHs of high-redshift galaxies, another interesting issue is the 
significant disagreement regarding the evolution of the specific SFR 
(SSFR; defined as SFR/$M_{*}$) at $z \gtrsim 2$ between the observational 
results and the prediction from galaxy formation models.
SAMs of galaxy formation predict that the SSFR increases with 
redshift beyond $z \gtrsim 2$ \citep[e.g.,][]{dut10}, while observational results 
show a plateau over a wide range of redshift, $2 \lesssim z \lesssim 7$ \citep[e.g.,][]{gon10}. 
Recently, there have been several works trying to resolve this tension. 
For examples, \citet{bou12} have re-calculated the SSFRs based on their new estimation 
of UV spectral slope ($\beta$) and dust attenuation, reporting higher values 
of SSFRs than previously estimated. 
Another improvement comes from the realization of the fact that the estimation of 
stellar mass can be affected by the contribution from the nebular emission 
\citep[e.g.,][]{deb12,gon12,sta13}. 
Even though the derived SSFR values show some discrepancies with each other, 
the results of these works commonly indicate that the stellar masses can be 
overestimated (thus SSFRs can be underestimated) without taking into account this effect. 
In this work, we explore this sharp contrast between model and observation, which 
imposes a challenge for our understanding of galaxy evolution. 
Our investigation of SSFR evolution is based on the consistent estimation of 
stellar masses and SFRs (thus, SSFRs also) from same SED-fitting (instead of 
estimating stellar mass and SFR separately, applying different assumptions), 
especially powered by more realistic assumption about SFHs of galaxies. 

In Section 2, we describe the observational data and sample selection. The details of  SED-fitting procedure and the results of this SED-fitting are explained in Section 3. 
In Section 4, we analyze the
results in more detail, especially focusing on the SFHs of high-redshift galaxies and 
the SSFR evolution at high redshift, and we analyze the ages of these galaxies in Section 5.
We summarize our results in Section 6. 
Throughout the paper, we adopt a flat $\rm{\Lambda}$CDM cosmology, with 
($\Omega_{m}, \Omega_{\Lambda}$) = (0.3,0.7), and 
$H_{0}$ = 70 $km$ $s^{-1}$ $Mpc^{-1}$. 
All magnitudes are given in AB magnitude system \citep{oke74}

\section{Observation and Galaxy Samples}
\subsection{Observation and Galaxy Catalog in the GOODS-S}

To understand the SFHs of high-redshift star-forming galaxies, we analyze the SEDs of 
high-redshift ($ 3 \lesssim z \lesssim 5$) LBGs observed in the CANDELS/GOODS-S field. 
The GOODS is a deep sky survey covering a combined area of $\sim$ 320 arcmin$^2$ in two 
fields --- the GOODS-N centered on the Hubble Deep Field-North (HDF-N) and the GOODS-S 
centered on the Chandra Deep Field-South (CDF-S). 
It carries an extensive wavelength coverage of the panchromatic photometric data from 
various facilities, including the $HST$/ACS (Advanced Camera for Surveys) and 
$Spitzer$/IRAC (Infrared Array Camera), supplemented with extensive spectroscopic 
follow-ups \citep{van08,van09,pop09}. 
Deep near-infrared (NIR) observation has been carried out in these two GOODS fields 
with $HST$/WFC3 (Wide Field Camera 3) as part of the Cosmic Assembly Near-infrared 
Deep Extragalactic Legacy Survey \citep[CANDELS,][]{gro11,koe11}.
The deep and extensive multi-band photometric data of the GOODS/CANDELS 
provide a unique and useful data set in the study of physical or stellar-population 
properties of high-redshift galaxies.

In our study, we use the photometric information from observed frame ultraviolet (UV) 
to mid-infrared (MIR) in fifteen broad bands --- $U$-band from the CTIO (Cerro Tololo 
Inter-American Observatory) and the VLT/VIMOS (VIsible Multi-Object Spectrograph), 
$B_{435}$, $V_{606}$, $i_{775}$, and $z_{850}$-bands from $HST$/ACS, F098M, 
F105W ($Y$), F125W ($J$), and F160W ($H$) from $HST$/WFC3, 
$Ks$-bands from VLT/ISAAC (Infrared Spectrometer and Array Camera), and 
$m_{3.6 \mu m}$, $m_{4.5 \mu m}$, $m_{5.8 \mu m}$, and $m_{8.0 \mu m}$ from 
Spitzer/IRAC. 
In selecting the high-redshift galaxies based on their color(s) or in analyzing their 
observed SEDs, it is very important to measure accurately in an unbiased way the 
colors or the SEDs from the observed photometric data which are typically from the 
instruments with different spatial resolutions and/or point spread functions (PSFs). 
Colors or SEDs with mismatched photometry can lead to unwanted biases in the 
inferences derived from the analysis of these SEDs or colors. In the case of the 
GOODS data, the resolution and PSF of the $HST$/ACS data are $\sim 0.03 \arcsec$ 
and $\sim 0.1 \arcsec$, respectively, while these for $Spitzer$/IRAC data are as 
large as $\sim 0.6 \arcsec$ and $\sim 2 \arcsec$, each. To construct reliable SEDs 
of galaxies, the matched photometry among various data from different facilities 
with different properties is performed using the TFIT \citep{lai07} photometric 
software by the CANDELS team. 
TFIT is a photometric package for matched photometry with the template-fitting technique. 
It uses the positional as well as the morphological information of high-resolution image 
($H$-band of $HST$/WFC3 in our case) as prior information to measure the corresponding 
fluxes at other bands with lower resolutions. 
Therefore, this package is a very useful tool in measuring colors or SEDs of galaxies 
from the observational data with different resolutions, especially for faint objects 
or in a crowded field. 
The details of the matched multi-band photometric catalogue of the 
CANDELS/GOODS-S field, which is used in this study, can be found in \citet{guo13}.

\subsection{Selection of Lyman Break Galaxies}

From the CANDELS/GOODS-S multi-band photometric catalogue, we select star-forming 
galaxies at redshifts $z \sim 3$, 4, and 5 using the spectral break at $\lambda \sim$ 912 
$\rm{\AA}$ in galaxy spectra (i.e., Lyman break) combined with blue color at longer 
rest-frame $UV$ wavelength. 
This Lyman break technique \citep{gia02} has been shown to select efficiently 
high-redshift, star-forming galaxies from optical photometric data 
sets \citep[e.g.][]{ste03}. 
We apply the same LBG selection criteria as in L09 and L10. 
The criteria are as follows:

\begin{equation} \label{Udrop1}
(U~-~B_{435})~\geq ~ 0.62 ~+~ 0.68 ~\times ~ (B_{435} ~-~ z_{850}) ~ \wedge
\end{equation}
\begin{equation} \label{Udrop2}
(U~-~B_{435})~\geq ~ 1.25 ~ \wedge
\end{equation}
\begin{equation} \label{Udrop3}
(B_{435}~-~z_{850}) ~\leq ~ 1.93
\end{equation}

for $U$-dropouts,

\begin{equation} \label{Bdrop1}
(B_{435} ~-~ V_{606}) ~> ~ 1.1 ~+~ (V_{606} ~-~ z_{850}) ~ \wedge
\end{equation}
\begin{equation} \label{Bdrop2}
(B_{435} ~-~ V_{606})~> ~ 1.1 ~ \wedge
\end{equation}
\begin{equation} \label{Bdrop3}
(V_{606}~-~z_{850}) ~< ~ 1.6
\end{equation}

for $B$-dropouts, and

\begin{eqnarray} \label{Vdrop1}
((V_{606} ~-~ i_{775}) ~> ~ 1.4667 ~+~ \nonumber \\
0.8889 ~\times ~ (i_{775} ~-~ z_{850})) ~ \vee \nonumber\\
((V_{606} ~-~ i_{775}) ~> ~2.0) ~\wedge
\end{eqnarray}
\begin{equation} \label{Vdrop2}
(V_{606} ~-~ i_{775}) ~> ~ 1.2 ~ \wedge
\end{equation}
\begin{equation} \label{Vdrop3}
(i_{775}~-~z_{850}) ~< ~ 1.3
\end{equation}

for $V$-dropouts.

Here, $\rm{\wedge}$ means logical `AND', and $\rm{\vee}$ is logical `OR'.
We also apply the ACS $z$-band magnitude cut to be $z_{850} \leq 26.6$, which
is the $S/N$=10 limit for extended sources in the GOODS-S.

From these color selection, 1879, 1066, and 165 $U$-, $B$-, and $V$-dropout 
candidates are selected. 
From these color selected candidates we exclude AGN- and stellar-candidates. 
AGN candidates are excluded based on: (1) $X$-ray data from Chandra 4Ms sample 
\citep{xue11}, (2) Radio sample of \citet{pad11}, or (3) IR selection using 
the method of \citet{don12} (provided to the CANDELS team by the courtesy of 
Jennifer Donley).
Based on these, 34, 23, and 1 objects are 
excluded as AGN candidates from $U$-, $B$-, and $V$-dropout sample, each.
Then, we remove 1, 3, and 1 stellar candidates from each dropout sample by applying
the criteria in stellarity parameter ($ \geq 0.9$) and in WFC3 $H$-band magnitude 
($H_{AB} \leq 24.0$). 

Among the remaining LBG candidates, 
35 ($U$-dropout), 52 ($B$-dropout), 15 ($V$-dropout) galaxies have spectroscopic data, 
and 3 $U$-dropout and 3 $V$-dropout candidate galaxies are excluded from each dropout 
sample due to their 
too low redshifts.
For the redshift cut, we apply minimum redshift requirement of $z$ = 1.8
($U$-dropout), 2.0 ($B$-dropout), and 3.0 ($V$-dropout). 
These minimum redshift cuts are based on the distribution of photometric redshifts 
of all LBG candidates. 
We have determined the minimum redshift for each dropout sample where the distribution 
shows a minimum or a significantly reduced tail toward low redshift. 
All 52 spectroscopic B-dropout candidates have redshift greater than 2.
The remaining $U$-, $B$-, and $V$-dropouts with spectroscopic redshifts are 32, 52,
and 12.

For the LBG candidates without spectroscopic redshift, we use photometric redshift 
derived by the CANDELS team based on a hierarchical Bayesian approach \citep{dah13}.
Among 1809, 988 and 148 $U$-,$B$-,and $V$-dropouts with only photometric redshift, 
we only include the galaxies with photometric redshift in the range of 
$1.8 \leq z_{phot} \leq 4.0$ in the case of $U$-dropout candidates, 
$2.0 \leq z_{phot} \leq 5.0$ in the case of $B$-dropout candidates, and 
$3.0 \leq z_{phot} \leq 6.0$ for the $V$-dropout candidates. 
Specifically, we exclude 906 ($U$-dropout), 220 ($B$-dropout), and 40 ($V$-dropout) 
galaxies from our sample for the following analysis based on their photometric redshift. 
These galaxies which are excluded due to their low photometric redshift have similar 
dropout- and rest-frame UV color with the remaining LBGs, but have, on average, redder 
color at longer wavelengths. 
This suggests that the majority of these galaxies are most likely low-redshift 
galaxies whose 4000 $\rm{\AA}$ breaks mimic the Lyman break of high-redshift LBGs. 
We can not completely rule out the possibility that some of these galaxies are actually 
high-redshift galaxies with redder color, probably due to higher attenuation.
However, excluding these galaxies would not significantly affect our conclusion unless 
these galaxies' star-formation properties are clearly different from the remaining LBG 
sample -- for example, deviating largely from the star-forming main sequence. 
Therefore, we keep our LBG sample conservative and more robust by excluding these 
potential low-redshift contamination.  

\begin{figure}
\plotone{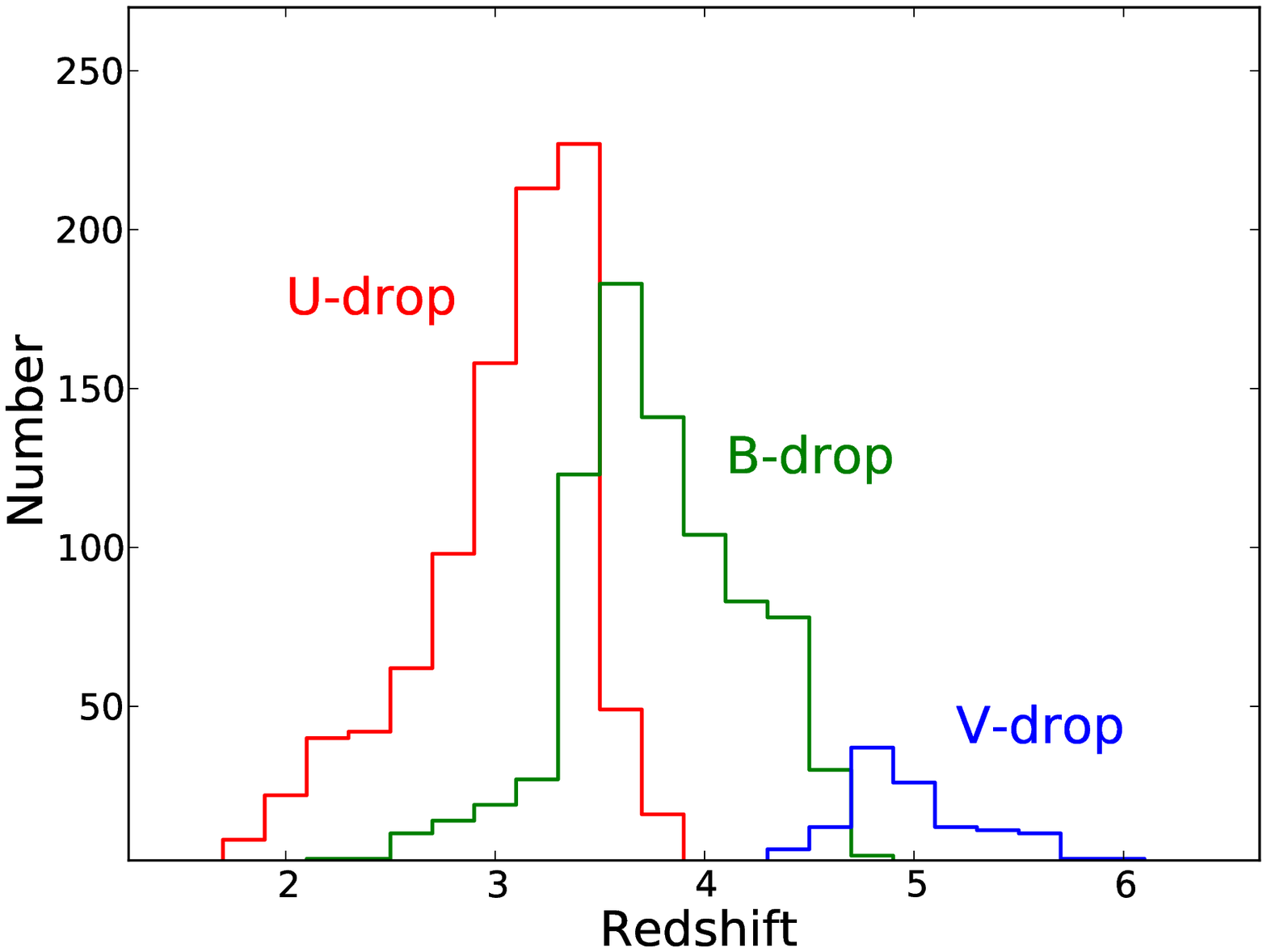}
\caption{Redshift distribution of the GOODS-S LBGs. The red solid line shows the redshift distribution of $U$-dropouts. The green solid line shows the distribution of $B$-dropouts. The blue line shows the distribution of $V$-dropouts. Photometric redshifts are used for galaxies for which spectroscopic redshifts are not available. \label{reddst}}
\end{figure}

Figure \ref{reddst} shows the redshift distributions of observed LBGs.
The red line shows the redshift distribution of $U$-dropouts, and the green line shows
the distribution of $B$-dropouts. The blue line is for $V$-dropouts.
For galaxies without spectroscopic redshifts, photometric redshifts are used in this figure.
The mean redshifts of each dropout samples are 3.0 ($U$-dropouts), 3.8 ($B$-dropouts), 
and 5.0 ($V$-dropouts). 
The over-density in this redshift distribution near $z \sim 3.4$ is also shown in 
photometric redshift distribution of total GOODS-S galaxies (Figure 20. in \citet{dah10}).

In Table. \ref{tab1},  we summarize the numbers of galaxies in each dropout sample.

\section{Spectral Energy Distribution Fitting}
\subsection{Fitting Procedure and Parameters}

To analyze the SEDs of GOODS-S LBGs and to constrain their physical properties, 
especially their SFHs, we perform the SED-fitting analysis using spectral templates 
from \citet[][hereafter, BC03]{bru03} stellar population synthesis models. 
We assume the Calzetti dust attenuation law \citep{cal00} for internal dust 
attenuation, and the \citet{mad95} extinction law for inter-galactic extinction due 
to the neutral hydrogen in the inter-galactic medium (IGM). 
We vary metallicity among three allowed values -- 0.2 $Z_{\odot}$, 0.4 $Z_{\odot}$, 
and 1.0 $Z_{\odot}$, where $Z_{\odot}$ is the solar metallicity. 
For the stellar initial mass function (IMF), we assume the \citet{cha03} form.

The main focus of this work is to investigate and constrain the SFHs of the observed 
high-redshift star-forming galaxies selected from their rest-frame $UV$ colors. 
Therefore, in this SED-fitting analysis, we assume two different forms of SFHs 
-- i.e., the exponentially-declining SFHs and the increasing ones.

In the case of exponentially declining SFHs, we vary the star-formation timescale 
parameter, $\tau$ -- which governs how fast the SFR declines with 
time -- from 0.2 Gyr to $\tau_{max}$, where $\tau_{max}$ = 2.0, 1.6 and 1.3 Gyr 
for $U$-,$B$-, and $V$-dropouts, respectively.
We vary the parameter $t$ -- which is the time since the onset of the star-formation 
-- from 50 Myr to $t_H$, where $t_H$ is the age of the Universe at each 
corresponding redshift.

The increasing SFHs used in this work have the same form as the `delayed
SFHs', introduced in L10, with the limited range of $\tau$. 
The delayed SFHs are parameterized as,

\begin{equation}
\Psi (t,\tau) \varpropto \frac{t}{\tau^{2}} e^{-t / \tau},
\end{equation} \label{delaysfr}

where $\Psi (t,\tau)$ is the instantaneous SFR.

This functional form provides the SFHs which initially increase with time and
then decline for large $t$ ($t > \tau$) after passing the peak at $t = \tau$. 
By setting the values of $\tau$ to be greater than the age of the universe at specific 
redshift, we can generate the SFHs in which the SFR keeps increasing until observed 
at the given redshift. 
To make a set of galaxy spectral templates with increasing SFHs, we vary
the values of $\tau$ as $\tau_{min} \leq \tau \leq 10.0$ Gyr. $\tau_{min}$ values are 
2.0 Gyr for $U$-dropouts, 1.5 Gyr for $B$-dropouts, and 1.0 Gyr for $V$-dropouts. 
This choice of $\tau$ range (to limit the range of the allowed SFHs) is 
motivated by the results of L10, which show that the 
predicted SFHs of high-redshift ($3 \lesssim z \lesssim 5$) star-forming galaxies increase 
with time from the analysis of SEDs of mock galaxies from the semi-analytic models 
(SAMs) of galaxy formation. 
We vary the $t$ within the same range as in the SED-fitting with exponentially 
declining SFHs, i.e., from $t=50$ Myr to the age of the Universe at given redshift.

In the SED-fitting procedure, we fix the redshift at the given spectroscopic 
redshift value whenever available, or at the photometric redshift otherwise. 
In Table. \ref{tab2}, we summarize the range of parameters used in SED-fitting 
for each SFH form.

\subsection{SED-fitting Results with Declining and Increasing SFHs}

As explained in the previous section, we perform SED-fittings with two 
different assumptions about the SFHs -- first, with the widely-used, 
exponentially declining SFHs and then with the increasing ones. 
The main aim of this experiment is to examine what form of SFH is more 
appropriate for the analysis of the SEDs of high-redshift LBGs -- in other 
words, to see which form of SFH is more representative to the actual SFHs 
of our observed LBGs.

As our first test on the SFHs of CANDELS/ 
GOODS-S LBGs, we compare the minimum 
(i.e., best-fit) $\chi^2$ values from the SED-fittings with two different assumptions 
about the SFHs.
The minimum $\chi^2$ value for each observed galaxy provides the measure of the match 
between the observed SED and the model galaxy SED templates.
Therefore, from this comparison of the minimum $\chi^2$ values, we can test which 
form of SFHs produces galaxy SEDs which are closer to the observed SEDs of the 
CANDELS/GOODS-S LBGs.

\begin{figure}
\plotone{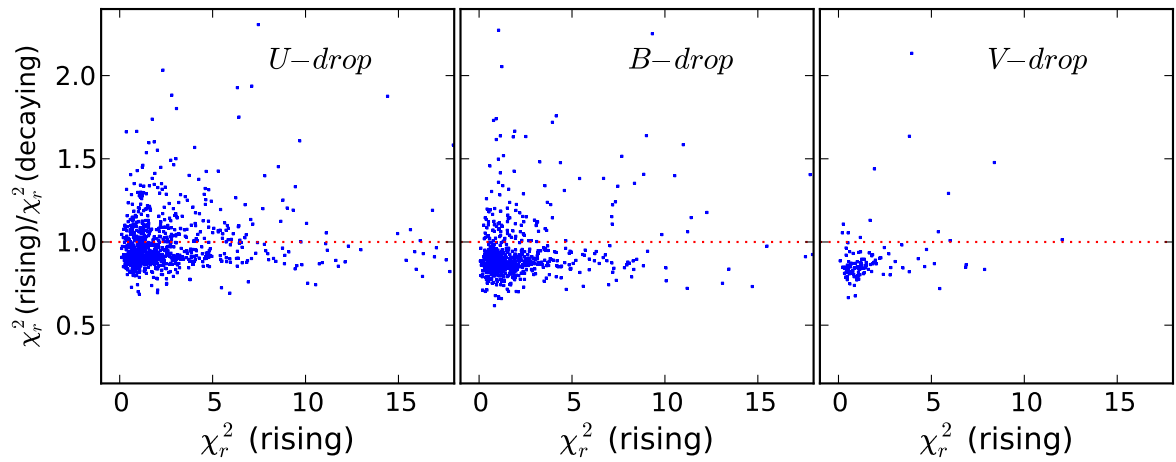}
\caption{Ratio between minimum $\chi^2$ values in the SED-fitting with the assumption of increasing SFHs and in the SED-fitting with declining SFHs. Each panel shows the $\chi^2$ ratios for $U$-, $B$-, and $V$-dropouts from left to right. \label{chisq}}
\end{figure}

In Figure \ref{chisq}, we show the ratio between the minimum $\chi^2$s in the SED-fitting
with the increasing SFHs and the declining SFHs, for $U$-, $B$-, and $V$-dropouts
(from left to right panel).
From this figure, we can see that the minimum $\chi^2$ values from the SED-fitting with
increasing SFHs are smaller than the values from the SED-fitting with
declining SFHs for the majority of the observed LBGs, especially for $V$-dropouts.
This means that the set of synthetic spectral templates with the increasing SFHs, on average, 
provide a better match to the SEDs of the observed GOODS-S LBGs.
In this figure, we can see that the minimum $\chi^2$ ratios ($y$-axis) are smaller than 
1 for most of $V$-dropout galaxies, while the fraction of galaxies with the ratio $\gtrsim 1$ 
progressively increases with decreasing redshift -- i.e., the $\chi^2$ difference from 
the SED-fitting with different SFH assumptions is more clearly seen at higher redshift.
The median values of $\chi^2$ ratio are 0.93, 0.88, and 0.86 for $U$-, $B$-, and 
$V$-dropouts, respectively.

While the comparison of minimum $\chi^2$ values with different SFHs favours the rising 
SFHs, it should be noted that the difference between the 
two minimum $\chi^2$ values is not significant for most individual galaxies 
-- the ratio values are clustered near 1 (shown as red dotted horizontal line in each 
panel).
Furthermore, while increasing SFHs appear to be favoured on average, we
have not explored other possibilities such as varying the dust attenuation law,
the IMF, or the IGM attenuation treatment.
Therefore, this $\chi^2$ comparison can only be considered as one of the supporting 
evidence favouring the rising SFHs, and we will provide further evidences in the 
following section.

\section{Analysis: Star-formation Histories of High-redshift Galaxies}
\subsection{Correlation between Stellar Masses and Star Formation Rates}

Previous studies have shown that there is a positive correlation between 
stellar masses and SFRs for high-redshift star-forming galaxies 
\citep{pap06,dad07,elb07,noe07,pan09}.
This locus of galaxies has been dubbed as the ``star-forming main sequence of galaxies".
This observed tight correlation indicates that galaxies' mass 
plays a crucial role in regulating their star formation activity.
The investigation of this correlation between galaxies' stellar masses and SFRs and
its evolution provides us an important clue to their formation histories
\citep[e.g.,][]{ren09,saw12}.

\begin{figure}
\plotone{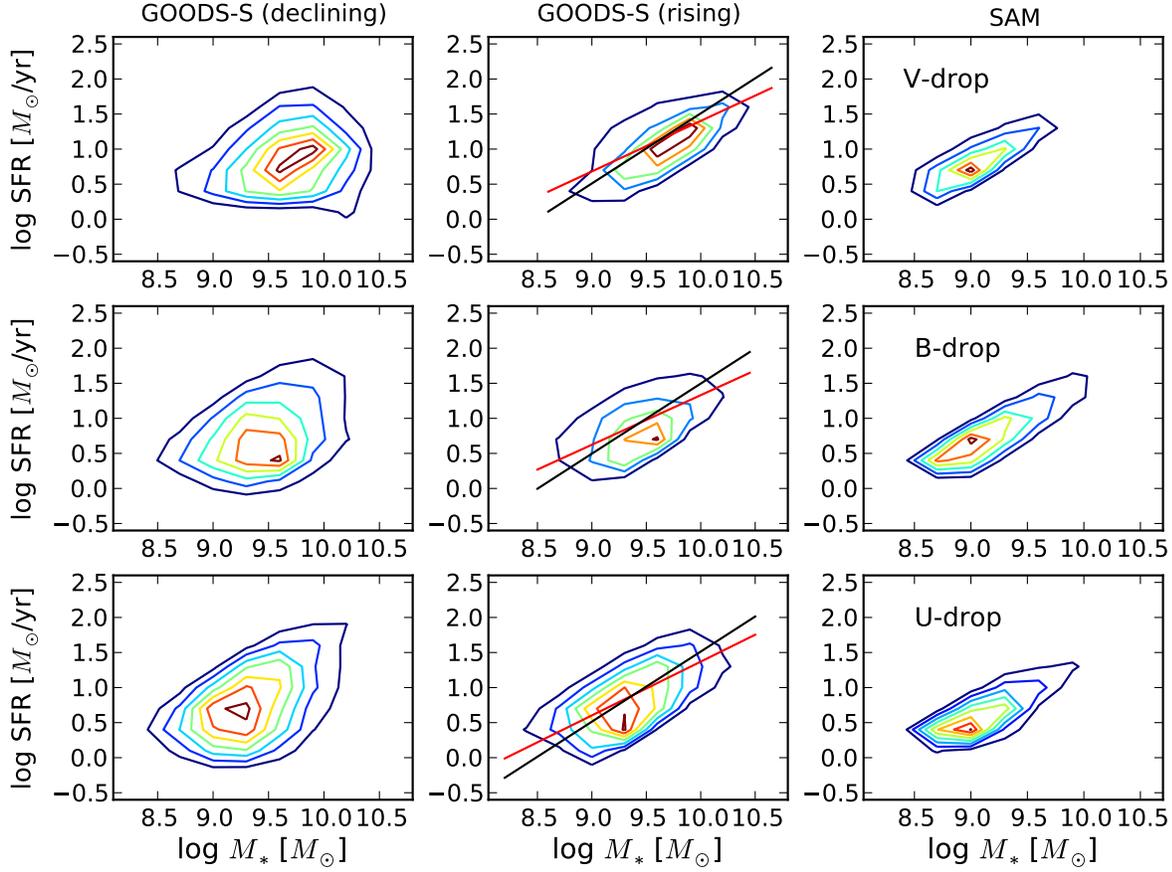}
\caption{{\bf (Left and Middle) :} Correlation between the stellar masses ($x$-axis) and the star-formation rates (SFR, $y$-axis) of observed Lyman-break galaxies in the GOODS-S. {\bf (Right) :} Same correlation for the model galaxies from the semi-analytic models (SAM) of galaxy formation. Left column shows the results from SED-fitting with the exponentially declining star-formation histories (SFHs)
while middle column is the results from SED-fitting with the increasing SFHs. In the middle panels, we show the best-fit linear correlation (black line in each panel) as well as the best-fit power-law correlation (red line). \label{mssfr}}
\end{figure}

Here, we use our LBG sample to explore the SF properties along this
star-forming main sequence at $z \geq 3$.
Figure. \ref{mssfr} shows the distributions of the GOODS-S LBGs as well as the mock
LBGs from semi-analytic models of galaxy formation in the stellar-mass--SFR plane.
Here, we use the CANDELS/GOODS-S mock galaxy catalogue.
This mock catalogue is generated by the CANDELS team combining the
Bolshoi N-body simulations \citep{kly11} and 
the semi-analytic models of \citet{som12}, which is an updated version of 
\citet{som01} and \citet{som08}.
This model (and previous versions of it) has been used in various studies of
galaxy properties \citep[e.g.][]{idz04,fon09,fon12,nie12,por12}, providing meaningful
insights into the galaxy evolution.

Our SAM assumes the standard $\Lambda$CDM universe and includes the theory of the growth
and collapse of fluctuations through gravitational instability. It implements the method
of dark matter halo merger tree construction of \citet{som99}. After this construction of
dark matter halo merger tree, analytic treatments for various physical processes,
including gas cooling, star formation, chemical evolution, mergers of galaxies, and
feedback effects from supernova and from AGN (active galactic nuclei), are applied.
For the detailed description about these analytic treatments, please refer above references.

The major update done in \citet{som12} is the treatment of the absorption and re-emission
of stellar light by the interstellar medium (ISM) dust. Briefly, dust extinction is
modelled based on \citet{cha00}, considering two components -- i.e, dust extinction
associated with the birth clouds around young star-forming regions, and extinction due to
the diffuse dust in the disc -- differently.
The wavelength dependence of the dust attenuation is modelled assuming the \citet{cal00}
starburst attenuation law. The absorbed energy is re-emitted in the infrared wavelength range.
The model uses dust emission templates to determine the dust emission SEDs.
For the detailed explanation about this update in the treatment of dust absorption and
emission, please refer \citet{som12}.
From this SAM mock catalogue, we select mock $U$-,$B$-, and $V$-dropouts, applying 
the same 
color selection criteria which are used to select our observed CANDELS/GOODS-S 
LBG samples (i.e., Equations (1)-(9)).

For this investigation, we focus on the observed LBGs with their best-fit $t$ values
not smaller than $100$ Myr.
The best-fit $t$ values of $t < 100$ Myr imply these galaxies might be caught during
their star-bursting event or during their very initial phase of formation.
By excluding these $starburst$-$candidate$ (i.e., galaxies with very small best-fit 
$t$s), we focus on the stellar-mass--SFR correlations
of continuously star-forming galaxies (i.e., star-formation main sequence galaxies) because
100 $Myr$ is a longer time scale than the typical dynamical time of high-redshift
galaxies \citep[e.g.][]{wuy11}.
The fractions of these $starburst$-$candidates$ are $13 \%$ among $U$-dropouts, 
$15 \%$ among $B$-dropouts, and $16 \%$ among $V$-dropouts when
we fit the observed SEDs with increasing SFHs.
The fractions are slightly higher in the case of SED-fitting with declining SFHs: 
$13 \%$ for $U$-dropouts, $19 \%$ for $B$-dropouts, and $23 \%$ for 
$V$-dropouts.
We discuss, in more detail, about these galaxies with the very small best-fit
$t$s in the following section.

In Figure \ref{mssfr}, we compare the correlation between stellar masses and
the SFRs of the observed LBGs to the predictions from SAM.
The first two columns are the mass-SFR correlation of GOODS-S LBGs, while the 
right column shows the same correlation for SAM LBGs.
The left column shows this correlation of observed LBGs when the stellar masses 
and SFRs are estimated through SED-fitting with exponentially declining 
SFHs assumed, while the middle column shows the results for the same 
observed LBGs, but from the SED-fitting with increasing SFHs.
In the left and middle columns, contours are drawn from the results of 1000 
Monte-Carlo runs of SED-fitting.
In each Monte-Carlo run, we first scatter the measured flux by the 
amount of random error.
We assume the Gaussian distribution of random error with the measured flux
error for each galaxy as the $\sigma$ in the Gaussian function.
We further scatter the flux assuming $5\%$ of systematic error.
For the flux uncertainty, we sum the measured flux error and $5 \%$ of the 
measured flux in quadrature.
The SFR is the average SFR over past 100 Myr (i.e., the mass of stars formed
in the past 100 million year divided by the same time interval) throughout this paper.
It is clear from this figure that the stellar masses and the SFRs of GOODS-S LBGs
show a tight positive correlation only when these properties are estimated through
SED-fitting with the rising SFH assumption.
When declining SFHs assumed, the SFRs are underestimated, leading to the larger 
scatter and lower normalization in $M_*$-SFR relation as shown in the left column 
of this figure. 
This underestimation of SFR with declining SFH assumption is consistent with the 
results of \citet{red12} -- left panel of their Figure 4. 
The stellar population properties -- such as stellar mass, SFR, and age -- derived 
from SED-fitting vary depending on the assumed form of SFH because different SFHs 
result in different mass-to-light ratios (or the relative fraction of stellar 
masses in old and young stellar populations). 
Exponentially declining SFHs assume that SFR was always higher in the past, 
so the resulting mass-to-light ratio is generally higher than the case of 
increasing SFHs, unless there is huge difference in ages. 
Therefore, the expected trend when declining SFHs are assumed is that 
the derived SFR values are smaller if the derived values of stellar mass are similar, 
or the derived stellar masses are larger if the derived SFRs are similar -- resulting 
in lower values of SSFR in either case, compared to the results with increasing SFHs. 

As shown in the right column of this figure, the SAM also predicts a tight, positive
correlation between the stellar mass and the SFR of LBGs.
If we fit the SEDs of mock LBGs assuming rising SFHs, we recover similarly tight 
correlations.
As shown in L09 and L10, if the SEDs of mock LBGs are fitted with exponentially 
declining SFHs, the SFR values are significantly underestimated and the amount 
of underestimation is nearly independent of stellar mass. 
Therefore, the $M_*$-SFR correlations would differ significantly from the 
intrinsic ones with larger scatter and lower normalization (similar behaviour 
shown in the left column of Figure \ref{mssfr}). 
While not conclusive, the similarity between the second and third columns
implies that galaxies, on average, have rising SFHs along the 
star-forming main-sequence at $z \geq 3$.
In this figure, we can see the relative dearth of massive galaxies in SAM 
compared to the observation for high-redshift galaxies, as also noticed 
in \citet{fon09}, \citet{mar09}, and \citet{san12}.
However, there are only a few galaxies in the high-mass ends, and this 
discrepancy between model and observation can arise due to the stellar mass 
errors that can scatter some galaxies to higher masses. 

In the case of SED-fitting with the rising SFHs, we estimate the best-fit
power-law correlations between the stellar masses and the SFRs at different redshifts,
parametrized as $log(SFR) = a \times log(M_*)+ b$. First, we find the best-fit correlation
by varying both of $a$ and $b$ (shown as red line in each panel of the middle column).
The best-fit values of ($a,b$) are (0.77,-6.30) for $U$-dropouts, (0.71,-5.74) 
for $B$-dropouts, and (0.72,-5.81) for $V$-dropouts.
The scatters from these best-fit correlations, measured as the standard deviation 
of the distances from these correlations, are 0.22, 0.23 and 0.19 dex for $U$-,$B$- and 
$V$-dropouts. 
From this best-fit estimation, we can see that the slope $a$ is close to unity.
This means that the specific SFR (SSFR) -- defined as the SFR per unit stellar mass
(SFR/$M_{*}$) -- at each redshift is nearly constant over the entire range of stellar mass.
This strong dependence of galaxies' star-formation activity on their stellar mass might
be the manifestation of the correlation between the star-formation activity and the
underling dark matter halo mass.

Next, we estimate the best-fit correlation, by fixing the slope $a$ as unity
(shown as black line in each panel of the middle column).
In this case, the best-fit $y$-intercept (in log-log space) $b$ for each dropout
sample is -8.49 ($U$-dropouts), -8.50 ($B$-dropouts), and -8.49 ($V$-dropouts).
Our result at $z \sim 3.8$ ($B$-dropouts) is in a good agreement with \citet{lee11},
where the average SFR is derived from rest-frame $UV$ luminosity of stacked
photometry for $z \sim 3.7$ galaxies.

\subsection{Galaxies with the Best-fit $t < 100$ Myr}

When investigating the correlation between stellar masses and SFRs of $3 \leq z \leq 5$
LBGs, we exclude the LBGs with small best-fit $t$s ($t < 100$ Myr).
These small best-fit $t$ values ($< 100$ Myr) can be spurious due to the effects of
photometric scatter \citep[e.g.][]{saw12}.
In that case, the dust reddening values (which are subject to the degeneracy with age)
as well as the SFR values (which are affected by the estimation of dust reddening) of
these galaxies from SED-fitting are also unreliable.
For this reason, we exclude these galaxies in the analysis of the correlation
between stellar masses and SFRs.

These galaxies have, on average, smaller stellar masses and higher SFRs than galaxies 
with $t_{best} \geq 100$ Myr at given stellar mass, 
thus forming an upper envelope in $M_*$-SFR correlation.
Therefore, if we include these galaxies, the normalization of the correlation would
increase and the slope would become slightly flatter than reported in the previous section.
However, the main conclusions of this work -- including the tight correlation between
the stellar masses and SFRs, the slow evolution of this correlation, and the
increasing SFHs of LBGs -- are not affected.

The best-fit $t$s with $t < 100$ Myr of these galaxies -- if considered as real --
imply that they are caught in their very early stage of formation or are experiencing
secondary burst events (with certain amount of hidden stellar masses).
To see how much stellar mass can be hidden in these galaxies, we fit their SEDs with two
components of stellar population.

For both of young and old stellar components, we
assume the delayed SFHs given in Equation (10).
Because the best-fit $t$s are smaller than 100 $Myr$
($\leq 90$ $Myr$) in the single-component fitting,
we allow the values of $t_y$ within 50
$Myr \leq t_{y} \leq 90$ $Myr$ for the young stellar
component, and vary the $t_o$ within 500
$Myr \leq t_{o} \leq t(z)$ for the old stellar
component.
Here, $t(z)$ is the age of the universe at given
redshift for each galaxy.
The parameter $\tau$ in Equation (10) is set to be
$\tau = 0.1$ $Gyr$ for the young component and $0.5$ $Gyr$ for the old component.
Because $t_y \leq 90$ $Myr$, the SFH of the young
component is the increasing one with this choice of $\tau$ value.
We assume that the both components have the same values of metallicity and 
dust attenuation.

From the results of this two-component fitting,
we see how many galaxies can have significant fraction of stellar
mass in their old component.
Among 110 ($U$-dropout), 127 ($B$-dropout), and 19 ($V$-dropout) galaxies with $t < 100$ $Myr$, 
18 ($16\%$), 17 ($13\%$), and 3 ($16\%$) galaxies in each dropout sample have their old 
mass fraction greater than $30 \%$.
This implies that majority of these galaxies are in the stages of their initial formation 
while a small fraction of these galaxies may be experiencing a starburst 
event with hidden old stellar mass.
Definite discrimination between these two possibilities is beyond the scope of this paper.

\subsection{$M_*$--SFR Correlations and Star Formation Histories}

In previous section, we have shown that the observed GOODS-S LBGs follow
a well-defined $M_*$--SFR correlation with near unity slope in the redshift range of 
$3 \leq z \leq 5$,
when we estimate these quantities with the increasing SFH assumption.
This tight correlation is a continued trend from 
lower redshifts \citep{dad07,elb07,noe07,pan09}.
\citet{ren09} investigated the star-formation histories of star-forming
galaxies from this observed correlation between the stellar masses and the
SFRs in the redshift range, $z \leq 2$, based on the results of \citet{pan09}.
\citet{pan09} found the best-fit evolution of the correlation between
the stellar masses and the SFRs of star-forming galaxies ($z\lesssim 2$)
as follows,

\begin{equation} \label{panssfr}
\rm{SFR} \simeq (\it{M_{*}}/10^{11}M_{\odot})(t/3.4 \times 10^9 \rm{yr})^{-2.5}(\it{M_{\odot}}\rm{yr}^{-1}).
\end{equation}

From this evolution of the observed correlation, \citet{ren09} re-parametrized
the evolution of the SFR of star-forming galaxies at $z \leq 2$ as,

\begin{equation} \label{renssfr}
\frac{\rm{SFR}(\it{t})}{\rm{SFR(2~Gyr)}} = 5.66 \times \frac{M(t)}{M(\rm{2~Gyr})} \times \it{t}^{-2.5},
\end{equation}
where $t$ is in billion years.
While the normalization of Equation (11) is fixed, we can adjust
the normalization, in the case of Equation (12), by changing the values of
SFR(2 Gyr) and $M_{*}$(2 Gyr).

In this work, we extend the redshift range of the investigation on this
correlation between galaxies' stellar masses and SFRs up to $z \sim 5$.
Based on the results of our work on the evolution of this correlation at
redshifts $3 \leq z \leq 5$, we now turn to the analysis of the expected
formation histories of our CANDELS/GOODS-S galaxies at this redshift range.

\begin{figure}
\plotone{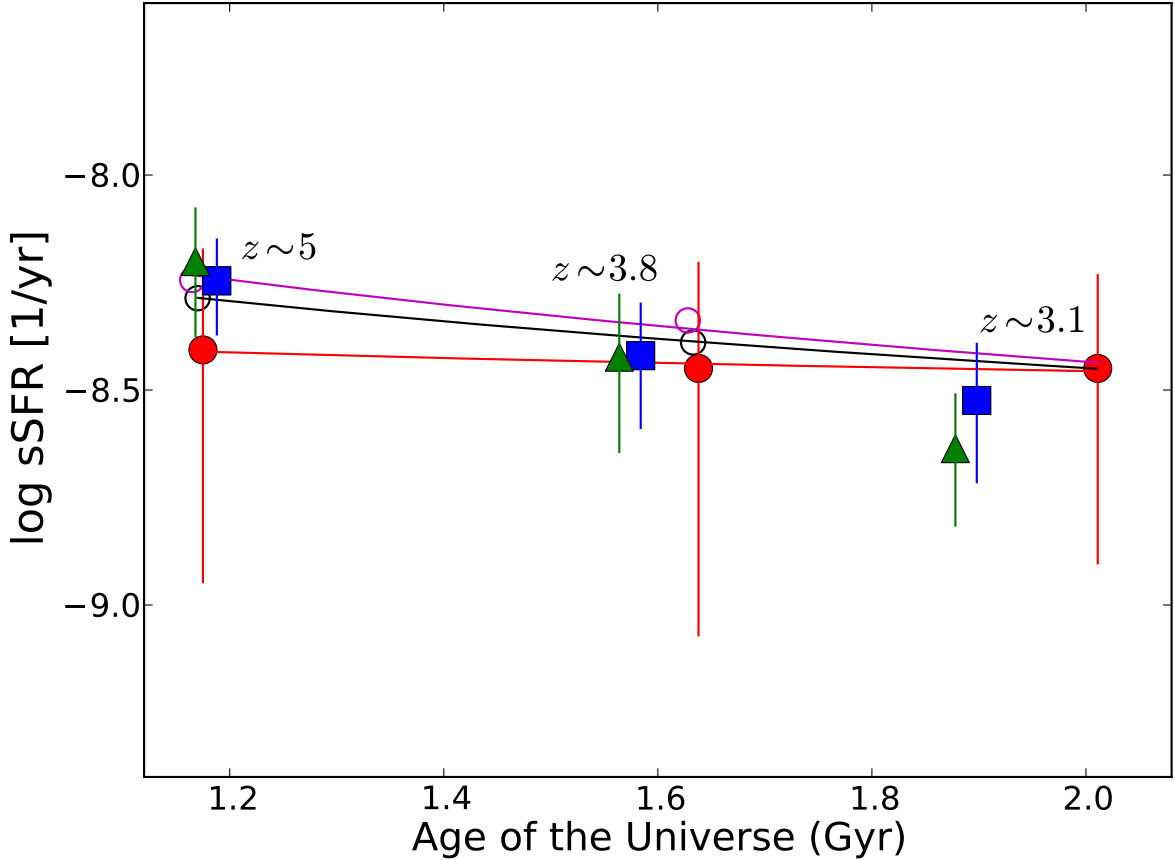}
\caption{Time evolution of mean specific SFR (SSFR) from the redshift $z\sim 5$ to 
$z \sim 3$.
Red circles with error bars show the mean SSFR values and uncertainties for each GOODS-S dropout sample.
The red solid curve is the best-fit parametrization of time-dependent SSFR evolution. 
The open circles are the mean SSFR values of the GOODS-S LBGs derived when we include the 
effects of nebular emission lines in the SED-fitting. 
These open circles are shifted slightly leftward for the visual clearance.
The black and purple circles are derived adopting the results of \citet{gon12} and 
of \citet{sta13}, respectively.
The black and purple solid lines are the best-fit curves of SSFR evolution for each case. 
Blue squares and error bars
show the mean SSFR evolution of SAM LBGs, while green triangles with error bars show 
the values derived through SED-fitting analysis of SAM LBGs. 
Blue and green symbols are shifted along the x-axis by +0.01 and -0.01 for the 
visual purpose, but these two symbols are at the same age points in x-axis. 
Time is given in Gyr. \label{ssfr}}
\end{figure}

For the LBGs with stellar mass in the range of 
$10^{9} \leq M_{*}/M_{\odot} \leq 10^{10}$, the mean SSFR is 
$3.55 \pm 2.30$ 1/Gyr for $U$-dropouts, and $3.55 \pm 2.70$ 1/Gyr 
and $3.92 \pm 2.79$ 1/Gyr for $B$- and $V$-dropouts, respectively.
In Figure \ref{ssfr}, we show the mean SSFR values at each
redshift as red circles with error bars and the best-fit SSFR evolution
as a red solid curve. 
The error bars show the standard deviation of SSFR. 
In this figure, we also show the SSFR evolution of model LBGs.
Blue squares with error bars show the intrinsic SSFR evolution of our SAM LBGs. 
Compared to the observed LBGs, our SAM LBGs show steeper SSFR evolution, and 
higher SSFR at the highest redshift bin ($z \sim 5$) by amounts of $\sim$ 0.16 dex. 
Green triangles and error bars show the SSFRs of SAM LBGs, which are derived 
from the SED-fitting with increasing SFHs assumed. 
The mean SSFR values derived from SED-fitting are similar with intrinsic values 
with slightly larger spread. 
Here, we do not show how the SSFR values would change if these values 
are derived from SED-fitting with exponentially decaying SFHs. 
However, it is obvious that the SSFR values would be much smaller if 
fitted with exponentially declining SFHs considering Figure 3 
(left column) as well as the results of L09 and L10. 
Figure 3 clearly shows that the best-fit SFR values of given 
stellar mass become much smaller if fitted with declining SFHs. 
In L09 and L10, we have shown that the SED-derived SFRs of mock LBGs from SAM are 
much smaller than the intrinsic values if they are derived from SED-fitting with 
declining SFHs. 

Recently, it has been suggested that the IRAC-bands (channels 1 and 2) 
might be affected at 
redshift $z \gtrsim 4$ due to the effects of nebular emission lines, such as 
H$\alpha$ or [\ion{O}{3}], resulting in overestimation of stellar 
mass and underestimation of SSFR \citep{deb12,gon12,sta13}.
With the inclusion of the effect of nebular lines, \citet{gon12} have 
found modest increases in the SSFR at $z \sim 4$ and 5, while the effect 
is larger at $z \sim 6$. 
This results have been acquired when they assume their $maximal$ emission 
line model where the equivalent width ($EW$) of H$\alpha$ increases 
with redshift as 

\begin{equation}\label{gonew}
EW(z) \sim 15.8 \times (1+z)^{1.52} \AA.
\end{equation} 

\citet{sta13} estimated the median excess of 0.27 magnitude at IRAC 
$3.6 \mu m$ for $3.8 < z < 5.0$ galaxies, which corresponds to 
$EW$ of 360--450 \AA. 
From this, they found the average reduction of stellar masses by 
factors of 1.1 and 1.3 at $z\sim 4$ and $z\sim 5$, respectively. 

We try to estimate and correct the effects of nebular emission 
lines in estimating SSFRs of our LBGs. 
First, we repeat the SED-fitting (with rising SFHs) correcting 
the flux values of the bands which can be affected by the 
existence of emission lines adopting the results of \citet{gon12} 
(their $maximal$ emission model). 
We get higher values of mean SSFRs -- $4.08 \pm 3.24$ 1/Gyr 
and $5.17 \pm 3.88$ 1/Gyr for $B$- and $V$-dropouts, respectively, 
for the galaxies with $10^{9} \leq M_{*}/M_{\odot} \leq 10^{10}$.
We show these values as well as the best-fit SSFR evolution 
as open black circles and black solid curve, respectively in Figure \ref{ssfr}. 
Then, we repeat the SED-fitting again adopting the results of 
\citet{sta13}. 
We show the corresponding results as open purple circles and purple 
curve in Figure \ref{ssfr}. 
The derived mean SSFRs are $4.59 \pm 3.66$ 1/Gyr for $B$-dropouts 
and $5.70 \pm 4.13$ 1/Gyr for $V$-dropouts. 
In either case, we get steeper slope of SSFR evolution which 
shows good agreement with the model prediction as well as similar 
SSFR values with the SAM LBGs at $z \sim 5$. 

If we parametrize the SSFR (with \citet{sta13} nebular correction) 
as $\rm{SSFR} = \it{c} \times t^{r}$,
the best-fit evolution of SSFR (i.e., purple curve in Figure~\ref{ssfr}) 
is given as

\begin{equation}
\rm{log ~SSFR} = -8.18 - 0.9 \rm{log} \it{t},
\end{equation}
where, $t$ is in Gyr. 

From this best fit, the evolution of SSFR within the redshift range of
$3 \leq z \leq 5$ is given as

\begin{equation} \label{messfr}
\rm{SFR} = 66.1 \times (\it{M_{*}}/10^{10} M_{\odot}) (t/1.0 \times 10^{9} \rm{yr})^{-0.9}.
\end{equation}

Equation (\ref{messfr}) has a smaller power in the time-dependent term than 
Equation (\ref{panssfr}), because the time-evolution of SSFR at $3 \leq z \leq 5$ 
is much slower than at $z \leq 2$.
From this best-fit SSFR evolution, we can connect the time evolution of SFR and
the time evolution of stellar mass as

\begin{equation} \label{mssfevol}
\frac{\rm{SFR}(\it{t})}{\rm{SFR}(\it{z}=5)} = (1.43 \times 10^8) \frac{M_{*}(t)}{M_{*}(z=5)} t^{-0.9},
\end{equation}

where $t$ is in year. 

The growth of stellar mass can be obtained by integrating Equation~(\ref{mssfevol}) 
as follows (by setting SFR = d$M_{*}$/d$t$),

\begin{equation} \label{msevol}
\frac{M_{*}(t)}{M_{*}(z=5)} = exp(-65.69) \times exp(8.15 t^{0.1}).
\end{equation}

\subsection{Comparison with Previous Work: SFH and Evolution of SSFR}

In the previous section, we analyze the history of star-formation activity and
of stellar mass assembly within the redshift range, $3 \lesssim z \lesssim 5$.
In this section, we compare our results with previous works in literature.

\citet{pap11} have derived the time evolution of the mean SFR at redshifts,
$3 \leq z \leq 8$ by analyzing the high-redshift galaxy samples with
constant comoving number density.
From this study, they show that the average SFR of high-redshift galaxies
increases as a power law with decreasing redshift (with best-fit power
$\alpha = 1.7$).

By combining Equations (\ref{mssfevol}) and (\ref{msevol}), we estimate 
the time evolution of SFR (i.e., the SFH) in the range of $3 \leq z \leq 5$.
By parametrizing the evolution of SFR as SFR($t$)/SFR($z=5$) = $(t/r)^{\alpha}$,
we find the best-fit power $\alpha = 5.85$ with the timescale parameter $r=1.16$ Gyr.
From this, we can confirm that the mean SFH of our GOODS-S LBGs steadily 
increases with time from $z \simeq 5$ to $z \simeq 3$ during about 1 billion 
years, which is in qualitative agreement with \citet{pap11}.
Our derived best-fit power is steeper than the value from \citet{pap11}.
When we try to find the power using only the data points at
$z$=3.1, 3.8, and 5.0 from \citet{pap11}, we find that the best-fit power is
$\alpha$=2.9.

Here, we derive the SFH from the best-fit estimation of SSFR evolution.
The SFR($t$) can also be derived more easily by finding the best-fit correlation
between the mean SFR and the cosmic time, $t$, at each redshift bin.
However, the mean SFR values can vary depending on the stellar
mass range of sampled galaxies because of the positive correlation between
stellar masses and SFRs.
The mean SSFR values are less subject to this problem because SFRs are
tightly correlated with stellar masses with the slope of near unity.
Therefore, SFH --- i.e., SFR($t$) --- derived from the evolution of SSFR is more
robust, independent of the sampled stellar-mass range.

\begin{figure}
\plotone{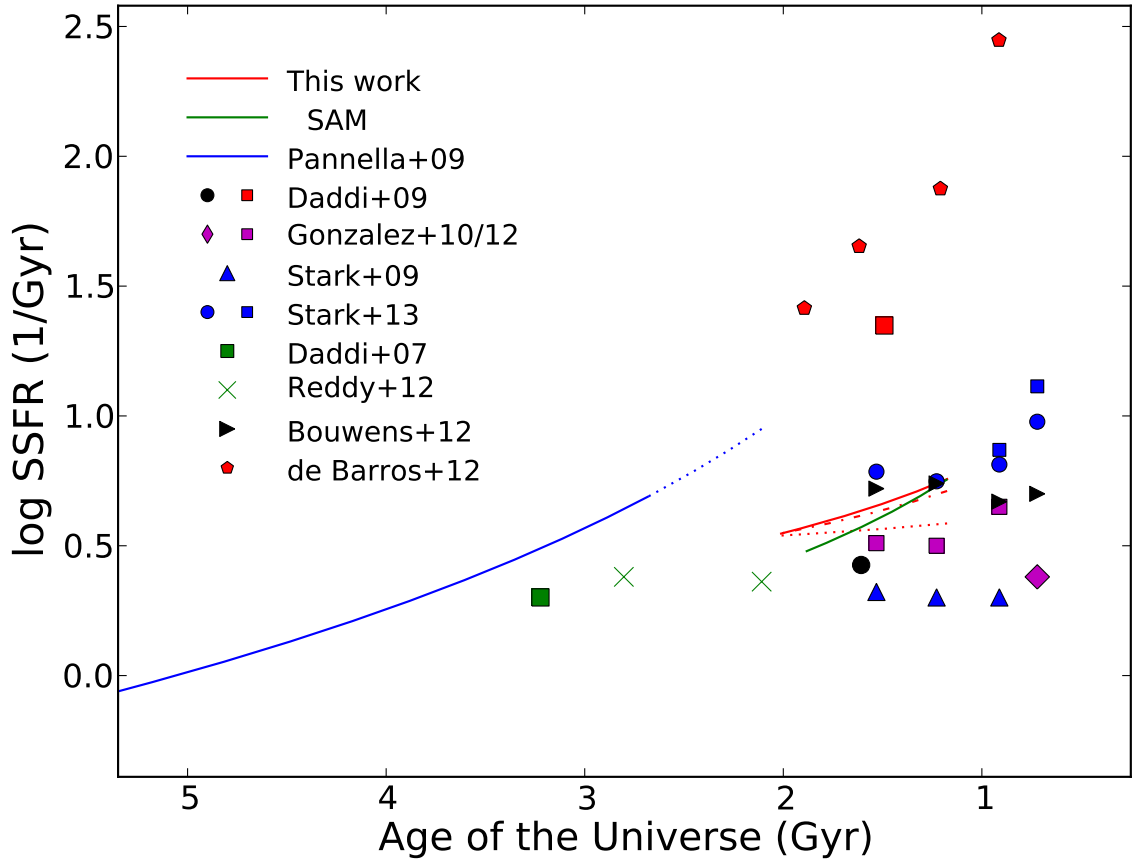}
\caption{Evolution of the SSFR of high-redshift galaxies. The estimation of SSFR evolution of 
$M_{*} \sim 5 \times 10^{9} M_{\odot}$ LBGs from the current work
($3 \leq z \leq 5$) is shown as the solid red curve. The blue curve shows the SSFR
evolution for $M_{*} \sim 3 \times 10^{10} M_{\odot}$ galaxies at 
$z \leq 3$ from \citet{pan09}. In the redshift range $2.4 \leq z \leq 3$, it is shown as dotted
curves as a visual reminder that the \citet{pan09} results are based on the observational data at
$z \lesssim 2.4$. The black circle and the red square are SSFR values of LBGs and of SMGs,
respectively, at $z \sim 4$ from \citet{dad09}. The purple diamond is from \citet{gon10}, 
and the blue triangles and the green square are the estimates made by \citet{gon10} based on the results of \citet{sta09} and \citet{dad07}, respectively. Green crosses are from \citet{red12}. 
Black trianles are the esitmation of \citet{bou12} based on their new estimation of UV slope. 
Blue circles and squares are the results of \citet{sta13} with fixed and evolving $H_{\alpha}$ 
EW, respectively. 
Purple squares and red circles are from \citet{gon12} and \citet{deb12}, each. 
Green curve is the prediction from our SAM. \label{ssfrevol}}
\end{figure}

Next, we compare our best-fit time evolution of SSFR with the one derived
by \citet{pan09} (or \citet{ren09}).
In Figure \ref{ssfrevol}, the blue curve shows the time evolution of SSFR 
for $M_{*} \sim 3 \times 10^{10} M_{\odot}$ star-forming galaxies 
derived by \citet{pan09} --- i.e., Equation (11) --- up to $z=3$.
Because Equation (11) is derived from the observational data up to
$z \leq 2.4$, we show the SSFR($t$) as the solid curve up to
$z \leq 2.4$ and as the dotted curve within the redshift range
$2.4 \leq z \leq 3$.
The time evolution of SSFR of our LBG sample at $3 \leq z \leq 5$ is shown
as the red curve in the same figure.
The solid and dashed red curves show the SSFR evolution when we adopt the 
nebular emission correction of \citet{sta13} and of \citet{gon12}, respectively. 
The dotted red curve is for the case when we do not apply any correction for 
the nebular emission.
It is clear that the SSFR($t$)s of \citet{pan09} and of ours 
show significant discrepancy at $z \sim 3$ --- SSFR($z=3$) from 
\citet{pan09} is much larger than our best-fit SSFR($z=3$).

For comparison, we show the mean SSFR value of $z \sim 4$ LBGs
($B$-dropouts) from \citet{dad09} as the black circle, which
shows a relatively good agreement with our results.
The mean SSFR of $z \sim 4$ submillimeter galaxies (SMGs) 
from the same authors is also shown in this figure as the red square.
It should be noted that our estimation of SSFRs at $z \geq 4$ is higher 
than previous results in literature. 
Purple diamond and blue triangles show the SSFR values between $z \sim 4$-7 
from \citet{gon10}. The SSFRs at $z \sim 4$, 5, and 6 are estimated by authors 
from the results of \citet{sta09}. 
Besides the lower values of SSFR, previous results also show a plateau between 
$z \sim 2$-7, which is contradictory to the theoretical prediction from SAMs.
Our results, on the other hand, indicate that the SSFR keeps increasing 
above $z \sim 2$, which is in good agreement with the prediction of our SAM, 
which is shown as the green curve.
Our improved agreement between observation and model prediction is 
encouraging because our result is based on (1) the consistent estimation of SFRs 
and stellar masses of our LBGs from same SED-fitting instead of estimating 
these quantities separately, and (2) appropriate assumption about the SFHs 
of high-redshift LBGs --- i.e., increasing SFHs.

Recently, \citet{bou12} reported higher dust extinction at 
$4 \lesssim z \lesssim 7$ based on their new measurement of UV-continuum 
slope, $\beta$. 
This leads to higher values of SSFR in this redshift range, shown as the black 
triangles in Figure \ref{ssfrevol}. 
Even though having higher values, the SSFR values still show little evolution 
with redshift. 

As mentioned above, \citet{deb12}, \citet{gon12}, and \citet{sta13} have included 
the effects of nebular emission in the estimation of SSFR through SED-fitting, 
and have reported higher SSFR values than previous results 
\citep[for example][]{gon10}, as well as increasing SSFRs with redshift in 
qualitative agreement with our results or the prediction from our 
SAM as shown in Figure~\ref{ssfrevol}. 

\section{Discussion: Ages of Star-forming Galaxies}

In spite of the relatively large uncertainties associated with the age 
estimation from SED fitting,
it is interesting to examine the age and age spread of galaxies at different
redshifts.
That is a consistency check for star-formation histories derived by looking
at stellar masses and SSFRs, and also provides some clues as to how well
synchronized galaxy formation was at early times.

\begin{figure}
\plotone{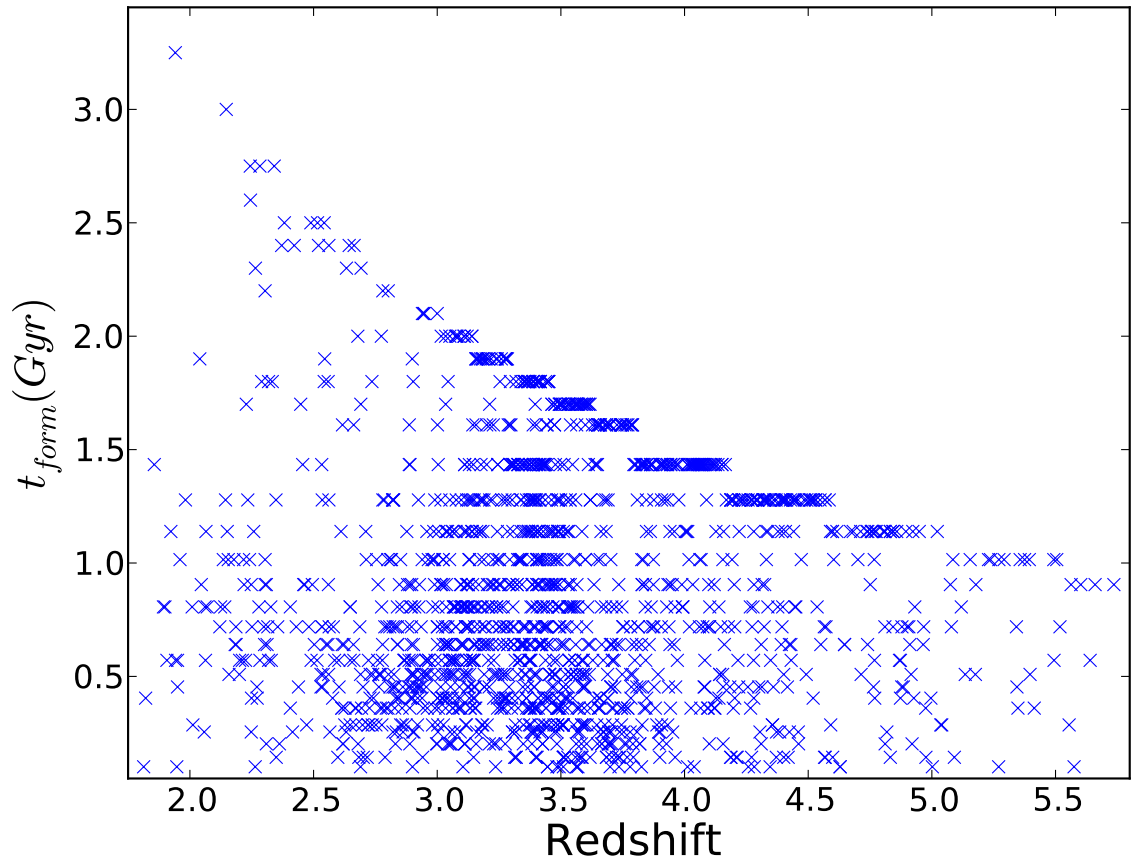}
\caption{Distribution of $t_f$ (in Gyr), the time since the onset of the star-formation, as a function of redshift. The upper envelope of this distribution corresponds to the age of the Universe at each redshift. \label{zt}}
\end{figure}

Figure \ref{zt} shows the distributions of $t_f$, the time since the formation of
each galaxy, as a function of redshift for the continuously star-forming, GOODS-S LBGs.
From this figure, we can see that many LBGs occupy the upper envelope of this
``redshift--$t_f$" distribution, which corresponds to the maximally possible $t_f$,
i.e., the age of the universe, at given redshift.
This suggests that many of the observed high-redshift LBGs initiated forming their stars very early.
Another noticeable feature in this redshift--$t_f$ distribution is that many galaxies
have relatively small values of $t_f$ ($< 800$ Myr) --- corresponding to formation
redshifts, $z_{f} \lesssim 6$ --- in the redshift range of $3.0 \lesssim z \lesssim 3.7$. 
Interestingly, this redshift range roughly coincides with the over-density shown
in redshift distribution of LBGs (Figure \ref{reddst}) as well as in the distribution 
of photometric redshift of total GOODS-S galaxies \citep{dah10}.
This feature indicates that not all star-forming galaxies were formed coevally even
at this high redshift range.

Next, we analyze the correlation between galaxies' stellar masses and their
stellar population mean ages.
This correlation also carries important information about the evolution of galaxies.
For example, for local galaxies, it is known that more massive galaxies host older
stellar populations, known as the archaeological downsizing.

\begin{figure}
\plotone{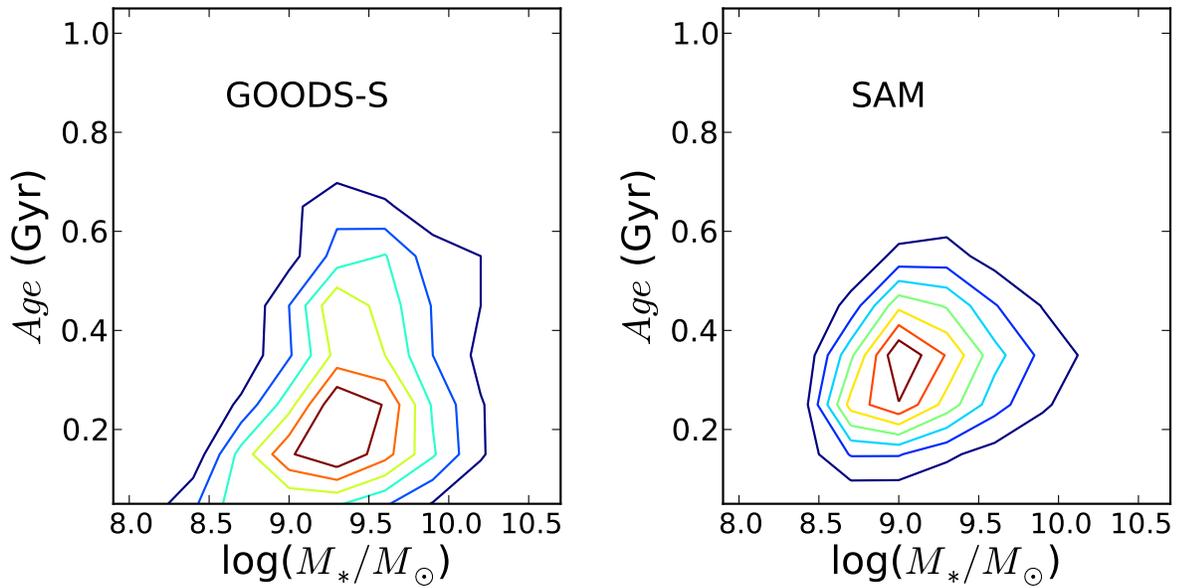}
\caption{The correlation between the stellar-masses and stellar population ages of high-redshift
galaxies. Here the ages are stellar-mass weighted mean ages in Gyr, and the stellar masses are in log scale with the unit of solar mass ($M_{\odot}$). {\bf (Left) :} The contour map of GOODS-S star-forming galaxies in the stellar-mass--age plane. {\bf (Right) :} The stellar-mass--age correlation of LBGs from semi-analytic models. \label{msage}}
\end{figure}

The left panel of Figure \ref{msage} shows the distributions of stellar population
mean ages of GOODS-S LBGs as a function of their stellar masses.
From this figure, we can see that there is a positive correlation between LBGs'
stellar masses and their stellar population ages -- in a sense that more
massive galaxies are on average older.
This positive correlation between galaxies' masses and ages is expected to arise
naturally if these galaxies form their stars in a continuous way with long duty
cycles with mass-dependent formation histories or star-formation timescales.
However, caution should be made because the correlation is relatively broad with
large scatter.
For example, galaxies with similar masses of $\sim 10^{9.3} M_{\odot}$ span
almost the entire range of stellar population ages allowed by the SED fitting.
This broad scatter in this correlation, while can be an effect of photometric
scatter or fitting anomaly, indicates that the formation or assembly histories
of these galaxies are governed not only by their mass but also by other
factors (like minor mergers or interactions).

In the right panel of Figure \ref{msage}, we show the $M_{*}$--age correlation
of SAM LBGs.
The stellar masses and ages of the SAM LBGs also show a positive correlation
with large scatter --- in a good agreement with the observed GOODS-S LBGs.
Besides the difference in the stellar mass range, mentioned in the Section 4.1,
the ages of the observed LBGs spans a wider range (toward old ages) than
the SAM LBGs.

\begin{figure}
\plotone{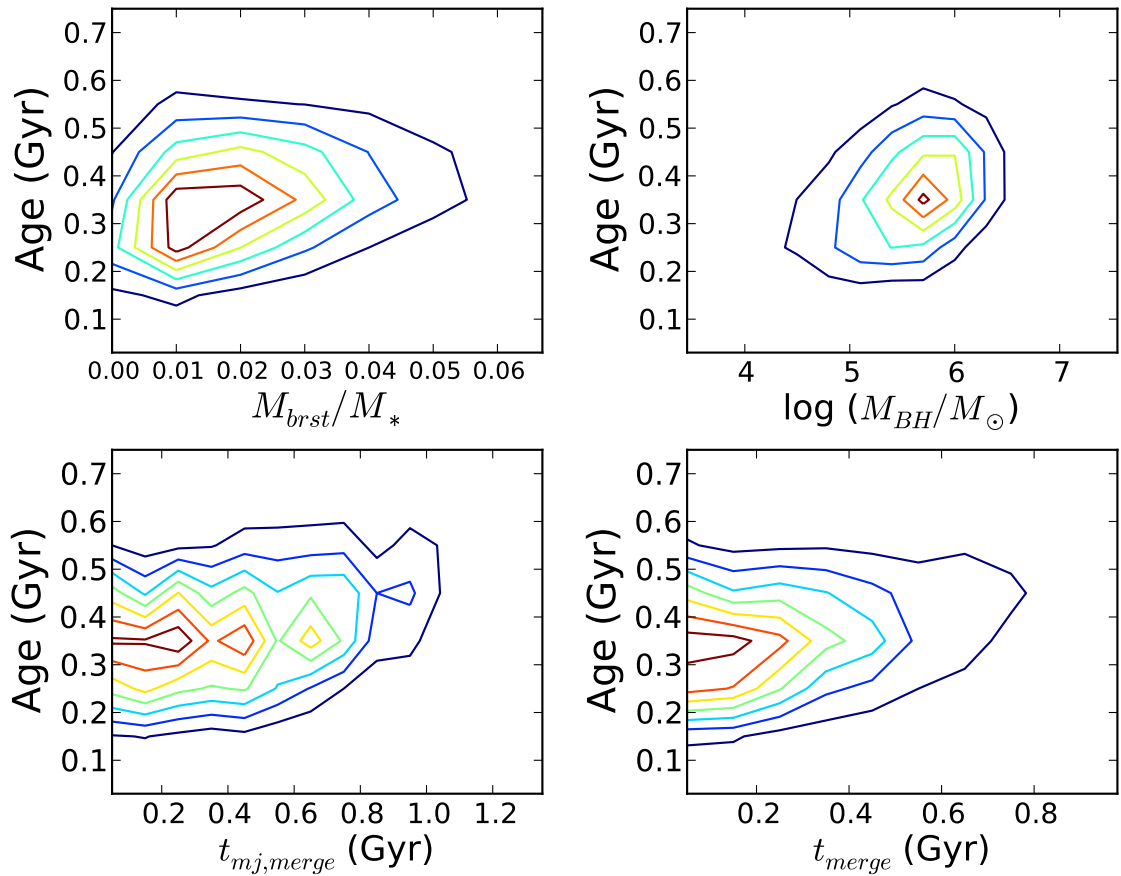}
\caption{Dependence of the stellar population mean ages on various galaxy properties for the 
$M_{*}/M_{\odot}  \sim 10^{9.2}$ mock LBGs from the SAM. In lower left panel, only the mock LBGs which experienced a major merger are shown. In all the other panels, the LBGs which experienced any merger (major or minor) are shown. $t_{merge}$ ($t_{mj,merge}$) is the elapsed time since the latest (major) merger event. $M_{brst}/M_{*}$ is the fraction of stellar mass formed during the merger-driven starburst. \label{samagedep}}
\end{figure}

To have an insight what can affect the age-spread for the galaxies with 
similar stellar mass, we check how the mean ages vary depending on various 
galaxy properties for the SAM LBGs with stellar masses in the range of
$10^{9.1} \leq M_{*}/M_{\odot} \leq 10^{9.3}$. 
Through this investigation, it has turned out that, among various properties, 
quantities related to the merger event show certain amount of correlation with the 
stellar population ages of galaxies. 
The upper panels of Figure \ref{samagedep} show that there are positive correlations 
between the mean age of LBGs and the burst mass fraction --- defined as the ratio of 
the stellar mass built during the merger-driven burst event ($M_{brst}$) to the total stellar 
mass ($M_{*}$) (left panel), as well as between the age and the black hole mass ($M_{BH}$) 
--- which grows through the merger event (right panel). 
In lower panels, we show the dependence of the galaxies' mean age on the time since the 
latest merger ($t_{merge}$, lower right panel) and since the latest major merger 
($t_{mj,merge}$, lower left panel). 
Here, only the galaxies which experienced merger(s) are shown, and in the lower left 
panel, we only show the galaxies which experienced a major merger. 
Among 6851 mock LBGs within this mass range, 79 $\%$ of galaxies experienced merger event(s) 
($49 \%$ for a major merger). 
When we compare the stellar population ages of mock LBGs with and without merger experience, 
they show clearer difference. 
The median value of the ages of galaxies which experienced merger events 
is 360 Myr, while it is 290 Myr for the galaxies without merger experience. 
This indicates that mergers or accretion of small galaxies have certain effects in shaping 
the galaxies SFHs, thus on the observed age-spread among galaxies with similar stellar mass.

\section{Conclusion}

In this work, we analyze the broadband photometric SEDs of observed GOODS-S
star-forming galaxies, selected by their rest-frame UV colors.
Through the detailed analysis of the SED-fitting results of these observed
galaxies with two different assumptions about their SFHs --- the exponentially
declining SFHs and the increasing SFHs, we examine the representative SFHs
of these high-redshift star-forming galaxies.
We also compare our results to the theoretical predictions from SAM for the
additional constraints on the SFHs of high-redshift star-forming galaxies.
Our main results are summarized as follows:

\begin{enumerate}

\item {The comparison of $\chi^2$ values from SED-fittings with different SFHs
indicates that the synthetic galaxy spectral templates with the increasing SFH
assumption provide better match to the observed SEDs of GOODS-S LBGs than the
templates with the exponentially declining SFH assumption do.}

\item {The stellar masses and the SFRs of observed GOODS-S LBGs show a tight
correlation, which is in good agreement with the predictions from SAM of galaxy
formation as well as with the observational results at lower redshift, when we
derive these quantities through the SED-fitting with increasing SFHs.
If we derive the masses and the SFRs through the SED-fitting with 
exponentially declining SFHs, the SFRs and stellar masses show much weaker 
correlation with large scatter.
This is thus an additional support for the plausibility of rising
SFHs for $z \geq 3$ LBGs.}

\item {From the observed tight correlation between galaxies' stellar masses 
and SFRs in the redshift range of $3 \leq z \leq 5$, we can also deduce the 
average SFH of these continuously star-forming galaxies in the given 
redshift range. From the evolution of the mean SSFR values, we infer the 
average SFH of galaxies at this redshift range as $\sim (t/1.16 \rm{Gyr})^{5.85}$ 
--- i.e., the average SFH of continuously star-forming galaxies, when the 
correction for the nebular emission included, increases steeply from 
$z \simeq 5$ to $z \simeq 3$ during about one billion years.}

\item {Our measured SSFR values show an increase with redshift from $z \sim 3$ to 
$z \sim 5$, indicating continuous increase from $z \sim 2$. While contrary to the previous 
results, this increasing trend is well consistent with recent results, such as 
\citet{deb12}, \citet{gon12}, and \citet{sta13}. The SSFR values, which are estimated 
in a consistent manner from SED-fitting, as well as its increasing trend are in good 
agreement with the prediction from the SAM.}

\item {While many of our observed LBGs lie on the upper envelope in the 
redshift-$t_f$ relation (i.e., have the largest possible $t_f$ at given redshift), 
there is a non-negligible fraction of galaxies whose $t_f$ values are much 
lower than the values at the envelope, especially, at redshifts 
$3.0 \lesssim z \lesssim 3.7$. This spread in $t_f$ implies that not all 
star-forming galaxies caught at redshifts $z \geq 3$ were formed coevally.}

\item {Our sample of continuously star-forming galaxies follows a positive 
correlation between their masses and stellar population ages, in a sense that 
more massive galaxies are, on average older, but with significant scatter 
--- similar with the prediction from the SAM. While this positive correlation implies 
that the formation histories of these star-forming galaxies vary depending 
on their masses, the broad scatter in this $M_{*}$--age correlation might 
be an indication that mass is not the only property which affects the 
formation histories of galaxies. The correlations between the ages and 
other properties of mock LBGs from the SAM suggest that mergers have certain 
effects on the age-spread among galaxies with given stellar mass.}

\end{enumerate}

From this SED-fitting analysis of the broadband SEDs of GOODS-S LBGs, we can 
confirm the important speculations of L09 and L10, which are drawn from the 
analysis of mock galaxies from SAMs: namely, (1) the assumed form of SFHs 
affect the estimation of stellar population parameters of galaxies from 
broadband SEDs, and (2) we should assume the increasing SFHs, not the 
exponentially declining ones, in the analysis of UV-selected, star-forming 
galaxies within the redshift range of $3 \lesssim z \lesssim 5$. 
Any bias arising in the fitting procedure due to incorrect assumption on 
the SFHs would propagate to the inferences on galaxy evolution drawn 
from the SED-fitting analysis.

The increasing SFHs inferred here from these SED-fitting arguments are qualitatively
consistent with the prediction from hierarchical models of galaxy formation in
$\Lambda$CDM \citep[e.g.][]{lee10,fin11}.
This smooth increase of SFR at high redshift is expected if the gas accretion 
rate closely traces the growth rate of halos, which grow as $\dot{M_h} \varpropto M_h$, 
and the SFR traces gas accretion rate.
Semi-analytic models and hydrodynamic simulations generically predict fairly smoothly 
rising SFHs at these redshifts \citep{lee10,fin07}. 
The generic rising form of the SFH is primarily due to the growth of structure 
\citep{her03}, but the {\it slope} of the SFH may be modulated by the overall efficiency 
of converting accreted gas into stars, which depends on the physics of star formation 
and stellar feedback, and may not be invariant with cosmic time.
Also, this continuously rising SFH provides a natural explanation for
the tight correlation between the stellar masses and the SFRs of star-forming galaxies
over a wide range of redshift, at least up to $z \leq 5$, as shown in this work ---
spanning more than 12 billion years, because galaxies would move along this relation
as they grow in time.
However, if high-redshift galaxies form their stars with decreasing SFRs with time,
this $M_{*}$--SFR correlation should become too broad with decreasing redshift,
unless all the LBGs were formed coevally --- which is not physically reasonable
and is also disfavored from the results in Section 5.
On the other hand, if the majority of star-forming galaxies form their stars in a
bursty way with short duty cycles, it is very hard to explain the formation of
this tight correlation.
In other words, why should the bursty star-formation activity care about the stellar
mass of host galaxy, which is mainly the result of past star-formation activity
(see also \citet{wuy11} and \citet{saw12} for the supporting evidence for the
continuous SFHs of high-redshift galaxies)?

Accurate determination of the shape of the SF main sequence and its intrinsic 
scatter will be very valuable to constrain the efficiency of star formation in 
different galaxies and environment. 
In fact, \citet{ren09} suggested that the rising slope of the SFH and the subsequent 
evolution are related --- the more gentle the rise, the more prolonged the overall 
SF activity. 
We speculate that if this is indeed the case, then the best predictor of the future 
evolution of SF of a galaxy is its position in the main sequence relative to other 
galaxies, assuming that the main sequence is accurately measured. 

The results of Section 5 --- broad distribution of formation times, and the
positive correlation between stellar masses and stellar population ages with large
scatter --- are broadly consistent with the expectations from the SAMs.
However, because the parameter $t$ or stellar population age is one of the most
poorly constrained parameters in SED-fitting \citep[e.g.][]{pap01,lee09,lee10,gua11},
it is still premature to draw firm conclusions.
While it is not easy to provide better constraints on the ages of these high-redshift
LBGs, more samples from the possible future surveys which are as deep as and wider
than the GOODS would be helpful by providing better statistics and enabling us to
study the dependence on interaction or environment.
Also, because our sample is selected based on the rest-frame UV, these may be
biased toward star-forming galaxies with moderate amount of dust extinction.
Inclusion of dusty star-forming galaxies, selected based on their rest-frame
optical \citep[e.g.,][]{guo12}, would provide more complete samples of high-redshift
star-forming galaxies, which will be helpful for understanding the formation
histories of high-redshift galaxies.

\acknowledgments

This work is based on observations taken by the CANDELS
Multi-Cycle Treasury Program with the NASA/ESA $HST$,
which is operated by the Association of Universities for Research
in Astronomy, Inc., under NASA contract NAS5-26555.
This work was supported by the National Research Foundation
of Korea (NRF) grant No. 2008-0060544, funded by the Korea
government (MSIP), and by the HST Program GO 12060 which
was provided by NASA through grants from the Space Telescope
Science Institute, which is operated by the Association
of Universities for Research in Astronomy, Inc., under NASA
contract NAS5-26555.

{\it Facilities:} \facility{HST (ACS, WFC3)}, \facility{Spitzer (IRAC)}, 
\facility{VLT (ISAAC, VIMOS)}, \facility{Blanco (MOSAIC II)}.

\clearpage
\onecolumn

\clearpage

\clearpage

\clearpage

\clearpage

\clearpage

\clearpage

\clearpage

\clearpage

\begin{deluxetable}{cccc}
\rotate
\tablecolumns{4} \tablewidth{0pc} \tablecaption{Numbers of Lyman Break Galaxies in the GOODS-S \label{tab1}}
\tablehead{ \colhead{Redshift}   &
\colhead{color-selected}   &
\colhead{excluding low $z_{spec}$}   &
\colhead{applying $z_{phot}$ cut\tablenotemark{b}  }} \startdata
3.0 & 1844 (35, 1809)\tablenotemark{a} & 1841 (32, 1809)\tablenotemark{a} & 935 (32, 903)\tablenotemark{a} \\
3.8 & 1040 (52, 988)\tablenotemark{a} & 1040 (52, 988)\tablenotemark{a} & 820 (52, 768)\tablenotemark{a} \\
5.0 & 163 (15, 148)\tablenotemark{a} & 160 (12, 148)\tablenotemark{a} & 120 (12, 108)\tablenotemark{a} \\
\enddata

\tablenotetext{a}{Values within the parenthesis are numbers of galaxies with and without spectroscopic redshift, respectively.}
\tablenotetext{b}{[Lower limit]: $z_{phot}$ = 1.8 ($U$-drop), 2.0 ($B$-drop), 3.0 ($V$-drop); [Upper limit]: $z_{phot}$ = 4.0 ($U$-drop), 5.0 ($B$-drop), 6.0 ($V$-drop)}

\end{deluxetable}

\clearpage

\begin{deluxetable}{cccccc}
\rotate
\tablecolumns{6} \tablewidth{0pc} \tablecaption{Fitting Parameter range in SED-fitting \label{tab2}}
\tablehead{ \colhead{SFH}   &
\colhead{$t$ (Gyr)}   &
\colhead{$\tau$ (Gyr)}   &
\colhead{Metallicity ($Z_{\odot}$)}   &
\colhead{Internal Dust Reddening}   &
\colhead{IGM Extinction}} \startdata
Exponentially Declining & 50 Myr - $t_H$\tablenotemark{a} & 0.2 Gyr - $\tau_{max}$\tablenotemark{b} & 
0.2, 0.4, 1.0 & Calzetti ($0.0 \leq E(B-V) \leq 0.9$) & Madau \\
Increasing & 50 Myr - $t_H$\tablenotemark{a} & $\tau_{min}$\tablenotemark{c} - 10.0 Gyr & 
0.2, 0.4, 1.0 & Calzetti ($0.0 \leq E(B-V) \leq 0.9$) & Madau \\
\enddata

\tablenotetext{a}{$t_H$ is the age of the Universe at the redshift of each galaxy.}
\tablenotetext{b}{$\tau_{max}$ = 2.0 Gyr ($U$-drop), 1.6 Gyr ($B$-drop), and 1.3 Gyr ($V$-drop)}
\tablenotetext{c}{$\tau_{min}$ = 2.0 Gyr ($U$-drop), 1.5 Gyr ($B$-drop), and 1.0 Gyr ($V$-drop)}

\end{deluxetable}


\begin{thebibliography}{}

\bibitem[Bouwens et al.(2012)]{bou12} Bouwens, R. J. et al. 2012, \apj, 754, 83
\bibitem[Bruzual \& Charlot(2003)]{bru03} Bruzual, G. \& Charlot, S. 2003, \mnras, 344, 1000, BC03
\bibitem[Calzetti et al.(2000)]{cal00} Calzetti, D., Armus, L., Bohlin, R. C.,
Kinney, A. L., Koornneef, J., \& Storchi-Bergmann, T. 2000, \apj, 533, 682
\bibitem[Chabrier(2003)]{cha03} Chabrier, G. 2003, \pasp, 115, 763
\bibitem[Charlot \& Fall(2000)]{cha00} Charlot, S. \& Fall, S. M. 2000, \apj, 539, 718
\bibitem[Conroy(2013)]{con13} Conroy, C. 2013, \araa, accepted, arXiv:1301.7095
\bibitem[Conroy et al.(2009)]{con09} Conroy, C., Gunn, J. E., \& White, M. 2009, \apj, 699, 486
\bibitem[Daddi et al.(2009)]{dad09} Daddi, E., Dannerbauer, H., Stern, D., Dickinson, M.,
Morrison, G., Elbaz, D., Giavalisco, M., Mancini, C., Pope, A., \& Spinrad, H.
2009, \apj, 694, 1517
\bibitem[Daddi et al.(2007)]{dad07} Daddi, E. et al. 2007, \apj, 670, 156
\bibitem[Dahlen et al.(2010)]{dah10} Dahlen, T. et al. 2010, \apj, 724, 425
\bibitem[Dahlen et al.(2013)]{dah13} Dahlen, T. et al. 2013, in preparation
\bibitem[de Barros et al.(2012)]{deb12} De Barros, S., Schaerer, D \& Stark, D. P. 2012,  
arXiv:1207.3663
\bibitem[Donley et al.(2012)]{don12} Donley, J. L. et al. 2012, \apj, 748,142
\bibitem[Dutton et al.(2010)]{dut10} Dutton, A. A., van den Bosch, F. C., \& Dekel, A. 2010, \mnras, 405, 1690
\bibitem[Elbaz et al.(2007)]{elb07} Elbaz, D. et al. 2007, \aap, 468, 33
\bibitem[Finlator et al.(2007)]{fin07} Finlator, K, Dav\'{e}, R. \& Oppenheimer, B. D. 2007, \mnras, 376, 1861
\bibitem[Finlator et al.(2011)]{fin11} Finlator, K., Oppenheimer, B. D. \& Dav\'{e}, R. 2011, \mnras, 410, 1703 
\bibitem[Fontanot et al.(2012)]{fon12} Fontanot, F., Cristiani, S., Santini, P., Fontana, A., Grazian, A., \& Somerville, R. S. 2012, \mnras, 421, 241
\bibitem[Fontanot et al.(2009)]{fon09} Fontanot, F., De Lucia, G., Monaco, P., Somerville, R. S., \& Santini, P. 2009, \mnras, 397, 1776
\bibitem[Giavalisco(2002)]{gia02} Giavalisco, M. 2002, \araa, 40, 579
\bibitem[Gonz\'{a}lez et al.(2014)]{gon12} Gonz\'{a}lez, V., Bouwens, R. J., 
Illingworth, G., Labb\'{e}, I., Oesch, P., Franx, M., \& Magee, D. 2014, \apj, 781, 34
\bibitem[Gonz\'{a}lez et al.(2010)]{gon10} Gonz\'{a}lez, V., Labb\'{e}, I., Bouwens, R. J., Illingworth, G., Franx, M., Kriek, M., \& Brammer, G. B. 2010, \apj, 713, 115 
\bibitem[Grogin et al.(2011)]{gro11} Grogin, N. A. et al. 2011, \apjs, 197, 35
\bibitem[Guaita et al.(2011)]{gua11} Guaita, L. et al. 2011, \apj, 733, 114
\bibitem[Guo et al.(2012)]{guo12} Guo, Y. et al. 2012, \apj, 749, 149
\bibitem[Guo et al.(2013)]{guo13} Guo, Y. et al. 2013, \apjs, 207, 24
\bibitem[Hernquist \& Springel(2003)]{her03} Hernquist, L. \& Springel, V. 2003, \mnras, 341, 1253
\bibitem[Idzi et al.(2004)]{idz04} Idzi, R., Somerville, R., Papovich, C., Ferguson, H. C., Giavalisco, M., Kretchmer, C., \& Lotz, J. 2004, \apj, 600, L115
\bibitem[Klypin et al.(2011)]{kly11} Klypin, A. A., Trujillo-Gomez, S. \& Joel, P. 2011, 
\apj, 740, 102
\bibitem[Koekemoer et al.(2011)]{koe11} Koekemoer, A. M. et al. 2011, \apjs, 197, 36
\bibitem[Laidler et al.(2007)]{lai07} Laidler, V. G. et al. 2007, \pasp, 119, 1325
\bibitem[Lee et al.(2011)]{lee11} Lee, K.-S. et al. 2011, \apj, 733, 99
\bibitem[Lee et al.(2010)]{lee10} Lee, S.-K., Ferguson, H. C., Somerville, R. S.,
Wiklind, T., \& Giavalisco, M. 2010, \apj, 725, 1644 (L10)
\bibitem[Lee et al.(2009)]{lee09} Lee, S.-K., Idzi, R., Ferguson, H. C., Somerville, R. S.,
Wiklind, T., \& Giavalisco, M. 2009, \apjs, 184, 100 (L09)
\bibitem[Madau(1995)]{mad95} Madau, P. 1995, \apj, 441, 18
\bibitem[Maraston et al.(2010)]{mar10} Maraston, C., Pforr, J., Renzini, A., Daddi, E., Dickinson, M., Cimatti, A., \& Tonini, C. 2010, \mnras, 407, 830
\bibitem[Maraston et al.(2006)]{mar06} Maraston, C. et al. 2006, \apj, 652, 85
\bibitem[Marchesini et al.(2009)]{mar09} Marchesini, D. et al. 2009, \apj, 701, 1765
\bibitem[Menci et al.(2006)]{men06} Menci, N., Fontana, A., Giallongo, E., Grazian, A., \& Salimbeni, S. 2006, \apj, 647, 753
\bibitem[Nagamine et al.(2005)]{nag05} Nagamine, K., Cen, R., Hernquist, L., Ostriker, J. P., \& Springel, V. 2005, \apj, 618, 23
\bibitem[Niemi et al.(2012)]{nie12} Niemi, S.-M., Somerville, R. S., Ferguson, H. C., Huang, K.-H., Lotz, J., \& Koekemoer, A. M. 2012, \mnras, 421, 1539
\bibitem[Night et al.(2006)]{nig06} Night, C., Nagamine, K., Springel, V., \& Hernquist, L. 2006, \mnras, 366, 705
\bibitem[Noeske et al.(2007)]{noe07} Noeske, K. G. et al. 2007, \apj, 660, L43
\bibitem[Oke(1974)]{oke74} Oke, J. B. 1974, \apjs, 27, 21
\bibitem[Padovani et al.(2011)]{pad11} Padovani, P., Miller, N., Kellermann, K. I., 
Mainieri, V., Rosati, P., Tozzi, P. 2011, \apj, 740, 20
\bibitem[Pannella et al.(2009)]{pan09} Pannella, M. et al. 2009, \apj, 698, L116
\bibitem[Papovich et al.(2001)]{pap01} Papovich, C., Dickinson, M., \& Ferguson, H. C., 2001, \apj, 559, 620
\bibitem[Papovich et al.(2011)]{pap11} Papovich, C., Finkelstein, S. L., Ferguson, H. C., Lotz, J. M., \& Giavalisco, M. 2011, \mnras, 412, 1123
\bibitem[Papovich et al.(2006)]{pap06} Papovich, C. et al. 2006, \apj, 640, 92
\bibitem[Popesso et al. (2009)]{pop09} Popesso, P. et al. 2009, \aap, 494, 443
\bibitem[Porter et al.(2012)]{por12} Porter, L. A., Somerville, R. S., Croton, D. J., Covington, M. D., Graves, G. J., Faber, S. M., \& Primack, J. R. 2012, \mnras, submitted, arXiv:1201.5918
\bibitem[Reddy et al.(2012)]{red12} Reddy, N. A., Pettini, M., Steidel, C. C., Shapley, A. E., Erb, D. K., \& Law, D. R. 2012, \apj, 754, 25
\bibitem[Renzini(2009)]{ren09} Renzini, A. 2009, \mnras, 398, L58
\bibitem[Salim et al.(2005)]{sal05} Salim, S. et al. 2005, \apj, 619, L39
\bibitem[Santini et al.(2012)]{san12} Santini, P. et al. 2012, \aap, 538, A33
\bibitem[Sawicki(2012)]{saw12} Sawicki, M. 2012, \mnras, 421, 2187
\bibitem[Sawicki \& Yee(1998)]{saw98} Sawicki, M. \& Yee, H. K. C. 1998, \aj, 115, 1329
\bibitem[Shapley et al.(2001)]{sha01} Shapley, A. E., Steidel, C. C., Adelberger, K. L., Dickinson, M., Giavalisco, M., \& Pettini, M. 2001, \apj, 562, 95
\bibitem[Somerville et al.(2012)]{som12} Somerville, R. S., Gilmore, R. C., Primack, J. R., \& Dominguez, A. 2012, \mnras, 423, 1992
\bibitem[Somerville et al.(2008)]{som08} Somerville, R. S., Hopkins, P. F., Cox, T. J., Robertson, B. E., \& Hernquist, L. 2008, \mnras, 391, 481
\bibitem[Somerville \& Kolatt(1999)]{som99} Somerville, R. S. \& Kolatt, T. S. 1999, \mnras, 305, 1
\bibitem[Somerville et al.(2001)]{som01} Somerville, R. S., Primack, J. R. \& Faber, S. M. 2001, \mnras, 320, 504
\bibitem[Stark et al.(2009)]{sta09} Stark, D. P., Ellis, R. S., Bunker, A., Bundy, K., Targett, T., Benson, A., \& Lacy, M. 2009, \apj, 697, 1493
\bibitem[Stark et al.(2013)]{sta13} Stark, D. P., Schenker, M. A., Ellis, R., 
Robertson, B., McLure, R., \& Dunlop, J. 2013, \apj, 763, 129
\bibitem[Steidel et al.(2003)]{ste03} Steidel, C. C., Adelberger, K. L., Shapley, A. E., Petini, M., Dickinson, M., \& Giavalisco, M. 2003, \apj, 592, 728
\bibitem[Vanzella et al.(2008)]{van08} Vanzella, E. et al. 2008, \aap, 478, 83
\bibitem[Vanzella et al.(2009)]{van09} Vanzella, E. et al. 2009, \apj, 695, 1163
\bibitem[Wuyts et al.(2011)]{wuy11} Wuyts, S. et al. 2011, \apj, 738, 106
\bibitem[Xue et al.(2011)]{xue11} Xue, Y. Q. et al. 2011, \apjs, 195, 10

\end{thebibliography}
\end{document}